\documentclass[aps,preprint,nofootinbib,groupedaddress,pra,showpacs]{revtex4}

\usepackage{graphics}
\usepackage{graphicx}
\usepackage{amssymb}
\usepackage{amsmath}
\usepackage{amsfonts}

\begin{document}

\title{Study of Simulation Method of Time Evolution in Rigged QED}

\author{Kazuhide Ichikawa}
\author{Masahiro Fukuda}
\author{Akitomo Tachibana} 
\email{akitomo@scl.kyoto-u.ac.jp}
\affiliation{Department of Micro Engineering, Kyoto University, Kyoto 606-8501, Japan}

\date{\today}

\begin{abstract}
We discuss how we formulate time evolution of physical quantities in the framework of the Rigged QED (Quantum Electrodynamics).
The Rigged QED is a theory which has been proposed to treat dynamics of electrons, photons and atomic nuclei in atomic and molecular systems
 in a quantum field theoretic way.
To solve the dynamics in the Rigged QED, we need different techniques from those developed for the conventional QED.
As a first step toward this issue, we propose a procedure to expand  the Dirac field operator, which represents electrons, by the
electron annihilation/creation operators and solutions of the Dirac equation for electrons in nuclear potential. 
Similarly, the Schr\"{o}dinger field operators, which represent atomic nuclei, are expanded by nucleus annihilation/creation operators.
Then we derive time evolution equations for these annihilation and creation operators
and discuss how time evolution of the operators for physical quantities can be calculated.
In the end, we propose a method to approximate the evolution equations of the operators by the evolution equations for 
the density matrices of electrons and atomic nuclei.
Under this approximation, we carry out numerical simulation of the time evolution of electron charge density of a hydrogen atom.
\end{abstract}




\maketitle

\section{Introduction}  \label{sec:introduction}

Recently, the technology for observation and manipulation
of quantum systems involving the interaction between light and matter is developed with increasing speed
and new experimental tools are becoming available.
Remarkable features of such developments include real-time observation \cite{Drescher2002,Krausz2009},
single-photon generation \cite{Kimble1998,Claudon2010,Buller2010,Aharonovich2011} and so on. 
Then, as for the theoretical side, it would be desirable to develop a formalism to simulate dynamical phenomena
observed by such experiments 
treating light and matter and the interaction between them as accurately as possible. 
That is, by Quantum Electrodynamics (QED).

QED is probably the most successful fundamental physical theory we have.
It is the theory of electrically charged particles and photons and interactions between them,
taking the form of quantum theory of fields. 
Its accurate predictions include the Lamb shift of the energy levels of the hydrogen atom and 
the anomalous magnetic moment of the electron. 
In addition to such stationary properties, it can also predict dynamical properties like 
the cross section for the electron-positron scattering (Bhabha scattering) to a great accuracy. 
The successes are due to the development of the method to compute the QED interaction 
as a perturbation. The success of the perturbative approach, in turn, owes much to the capability 
of preparing the asymptotic states as the unperturbed states (and of course to the renormalizability). 
It has been long since such formalism was established \cite{Peskin,Weinberg}. 

However, use of the asymptotic states and subsequently the $S$-matrix formulation has shortcomings 
when we would like to know the step-by-step time evolution of the quantum system employing QED.
This is because it just describes the transition between infinite past (``in-state") and infinite future (``out-state").
Although it works fine for calculations of cross sections of scattering processes or energy shifts of bound states,
it cannot be used for simulating the time evolution of the quantum system. 

Hence, so far, simulations of time evolution of the quantum system involving light and matter are
performed using the time-dependent Schr\"{o}dinger or Dirac 
equation with classical electromagnetic fields ({\it i.e.} semi-classical treatment)
\cite{Bauer2006, Iwasa2009, Mocken2008}
 or using a quantized photon field with the very much simplified matter part (so that the interaction between 
the photon and matter is introduced somewhat in an {\it ad hoc} manner) \cite{Loudon,Meystre,Haroche}.
It is true that these approximations are appropriate for a wide range of systems. 
For example, the former treatment is reasonable for a photon field produced by laser and the latter treatment 
for a system in cavity QED.
We consider, however, it is of great importance to develop a simulation method based on QED 
in the form of quantum field theory, and without recourse to the perturbative approach.\footnote{
In Refs.~\cite{Krekora2005,Krekora2006}, a formalism to simulate the dynamics of the quantum Dirac field 
has been developed but in the absence of the photon field.
}

Besides the non-perturbative time evolution scheme, what we wish to add to the standard QED framework 
is the atomic nucleus. This is because, contrary to the high-energy physics application, the existence of 
the atomic nuclei is essential for atomic and molecular science in which we are interested.
We would like to have a field theoretic QED formalism which is applicable to low energy phenomena
such as chemical reactions and photoionization process.
In this paper, as has been proposed in Ref.~\cite{Tachibana2001,Tachibana2003}, 
we include the atomic nucleus degree of freedom in the framework of the Rigged QED.

The Rigged QED \cite{Tachibana2001,Tachibana2003,Tachibana2010} is a theory which has been proposed to treat dynamics of charged particles and photons in atomic and molecular systems in a quantum field theoretic way.
In addition to the ordinary QED which is Lorentz invariant with Dirac (electron) field and U(1) gauge (photon) field, Schr\"odinger fields which represent atomic nuclei are added. 
In this way, dynamics of nuclei and their interaction with photons can be treated in a unified manner.
Incidentally, we here note on semantics of the word ``rigged". 
It has a nautical connotation as in the phrase ``fully rigged ship" and means ``equipped".\footnote{
Similar use of the word ``rigged" is found in a quantum mechanical context as ``rigged Hilbert space" 
which means Hilbert space equipped with distribution theory \cite{Gelfand}.
But it should be noted that the concept of the rigged Hilbert space is not used in the rigged QED.
}
Namely, the Rigged QED is ``QED equipped with Schr\"{o}dinger fields" as we just described above.

Including the atomic nucleus degrees of freedom as quantum fields is crucial. 
Since we would like to treat the interaction between an electron and an atomic nuclei or between two nuclei
as the one which is mediated by the quantized photon field (not by the classical electromagnetic field),
we need to express the atomic nuclei as quantum fields. 
Then, it is necessary to compute the interaction non-perturbatively due to the existence of the bound states. 
Although this is a difficult task to achieve,
in this way, we can go beyond the quantum mechanical treatment with perturbative QED corrections.
We believe such a theoretical technique opens up a way to study and predict new phenomena. 

To solve the dynamics of the Rigged QED in a non-perturbative manner, we need different techniques from those developed for the standard QED.
In this paper, as a first step toward this issue, we propose a procedure to expand  the Dirac field operator by solutions
of the Dirac equation for electrons in nuclear potential and derive time evolution equations for the electron annihilation and  creation operators.
Similarly, the Schr\"{o}dinger field operators are expanded by nucleus annihilation and creation operators.
Then we derive time evolution equations for these annihilation and creation operators
and discuss how time evolution of the operators for physical quantities can be calculated.
In the end, we propose a method to approximate the operator equations by $c$-number equations
and show some numerical results. 

Before ending this section, we note on our notations and conventions which will be used in this paper. 
They mostly follow those of Refs.~\cite{Tachibana2001,Tachibana2003,Tachibana2010}.
We use the Gaussian system of units. 
$c$ denotes the speed of light in vacuum, $\hbar$ the reduced Planck constant and 
$e$ the electron charge magnitude (so that $e$ is positive).
We put a hat to indicate a quantum operator to distinguish it from a $c$-number. 
A dagger as a superscript is used to express Hermite conjugate. 
A commutator is denoted by square brackets as $[A,B] \equiv AB - BA$
and an anti-commutator by curly brackets as $\{A,B\} \equiv AB + BA$.
We denote the spacetime coordinate as
$x=(x^\mu)=(x^0,x^i)= (ct, \vec{r})$ where the Greek letter runs from 0 to 3 and the Latin letter from 1 to 3. 
We adopt the convention that repeated Greek indices implies a summation over 0 to 3.
Other summations are explicitly written.
The transformation between contravariant and covariant vectors are done by the metric tensor
$g_{\mu\nu} = {\rm diag}(1,-1,-1,-1)=g^{\mu\nu}$.
The gamma matrices are denoted by $\gamma^\mu$.

This paper is organized as follows. 
In the next section, the framework of the Rigged QED which is relevant to the present study is reviewed briefly. 
In Sec.~\ref{sec:expansion}, we introduce annihilation and creation operators for each field operator and 
describe how we expand the field operators. 
In particular, Sec.~\ref{sec:expansion_photon} discusses how we treat the photon field operator 
in a non-perturbative manner following Ref.~\cite{Tachibana2003}.
In Sec.~\ref{sec:evolution}, we derive time evolution equations for these annihilation and creation operators
and discuss how time evolution of the operators for physical quantities can be calculated.
In Sec.~\ref{sec:approximation},  
we propose a method to approximate the evolution equations of the operators by the evolution equations for 
the density matrices of electrons and atomic nuclei.
Under this approximation, we carry out numerical simulation of the time evolution of electron charge density of a hydrogen atom.
The last section is devoted to our conclusion. 

\section{Rigged QED}  \label{sec:RQED}

In this section, we briefly review the general setting of Rigged QED \cite{Tachibana2001,Tachibana2003,Tachibana2010} by showing its Lagrangian and equations of motion for the field operators.

\subsection{Lagrangian}

First, we describe Rigged QED in terms of the Lagrangian. 
A part of the Rigged QED Lagrangian is the ordinary QED Lagrangian. 
The Lagrangian density operator of QED can be written as
\begin{eqnarray}
\hat{L}_{\rm QED}(x) = -\frac{1}{16\pi}\hat{F}_{\mu\nu}(x)\hat{F}^{\mu\nu}(x) 
+ \hat{L}_e \left( \left\{  \hat{\psi},   \hat{D}_{e\mu} \hat{\psi} \right\}; x \right),
\end{eqnarray}
where $\hat{F}_{\mu\nu}(x)$ is the electromagnetic field strength tensor which can be expressed by
the photon field $\hat{A}_\mu(x)$ as 
\begin{eqnarray}
\hat{F}_{\mu\nu}(x) 
&=&
\partial_\mu \hat{A}_\nu(x) - \partial_\nu \hat{A}_\mu(x). 
\end{eqnarray}
We note here that we adopt the Coulomb gauge $\vec{\nabla} \cdot \hat{\vec{A}}(x) = 0$ in this work.
$\hat{L}_e$ is the Lagrangian density operator for the electron  
\begin{eqnarray}
\hat{L}_e \left( \left\{  \hat{\psi},   \hat{D}_{e\mu} \hat{\psi} \right\}; x \right) 
=
c \hat{\bar{\psi}}(x) \left( i\hbar \gamma^\mu \hat{D}_{e\mu}(x) -m_e c   \right) \hat{\psi}(x),
\end{eqnarray}
where $m_e$ is the electron mass and the Dirac field operator $\hat{\psi}(x)$ represents the electron (and positron).
The operator with a bar on top is defined by $\hat{\bar{\psi}}(x)  \equiv \hat{\psi}^\dagger(x) \gamma^0$.
We denote the gauge covariant derivative for the electron as
\begin{eqnarray}
\hat{D}_{e\mu}(x) = \partial_\mu + i\frac{Z_e e}{\hbar c} \hat{A}_\mu(x), \quad Z_e = -1.
\end{eqnarray}

The Rigged QED Lagrangian is this QED Lagrangian ``rigged" with the Lagrangian of the atomic nuclei which are
represented by Schr\"{o}dinger fields. 
We denote the Schr\"{o}dinger field operator for the nucleus $a$ by $\hat{\chi}_a(x)$.
This satisfies commutation relations if $a$ is boson and anticommutation relations if $a$ is fermion (which in turn is determined by the nuclear spin of $a$).
The Lagrangian density operator for the atomic nucleus $a$ can be written as 
\begin{eqnarray}
\hat{L}_a \left( \left\{ \hat{\chi}_a, \hat{D}_{a 0} \hat{\chi}_a, \hat{\vec{D}}^2_a \hat{\chi}_a \right\}; x \right)
=
\hat{\chi}^\dagger_a(x) \left( i\hbar c \hat{D}_{a 0}(x) + \frac{\hbar^2}{2 m_a} \hat{\vec{D}}^2_a(x) \right) \hat{\chi}_a(x),
\end{eqnarray}
where $m_a$ is the mass of the nucleus $a$ and the gauge covariant derivative of $a$ is 
\begin{eqnarray}
\hat{D}_{a \mu}(x) = \partial_\mu + i\frac{Z_a e}{\hbar c} \hat{A}_\mu(x), 
\end{eqnarray}
where $Z_a$ is the $a$'s atomic number. 
Thus, when we have $N_n$ types of atomic nuclei in the system, 
\begin{eqnarray}
\hat{L}_{\rm Rigged QED}(x) &=& -\frac{1}{16\pi}\hat{F}_{\mu\nu}(x)\hat{F}^{\mu\nu}(x) 
+ \hat{L}_e \left( \left\{  \hat{\psi},   \hat{D}_{e\mu} \hat{\psi} \right\}; x \right) \nonumber \\
& &+ \sum_{a=1}^{N_n} \hat{L}_a \left( \left\{ \hat{\chi}_a, \hat{D}_{a 0} \hat{\chi}_a, \hat{\vec{D}}^2_a \hat{\chi}_a \right\}; x \right),
\end{eqnarray}
is the Rigged QED Lagrangian density operator. 
We note that this Lagrangian has $U(1)$ gauge symmetry but the Lorentz symmetry is broken by 
$\hat{L}_a$. This, however, will not be a problem since we are not going to solve the dynamics in 
a Lorentz covariant way.

\subsection{Equation of motion}

Here, we show the equations of motion for field operators,  $\hat{\psi}(x)$, $\hat{\chi}_a(x)$ and $\hat{A}_\mu(x)$, in Rigged QED. 
They are given by the principle of least action from the Lagrangian density operators introduced above. 
We also define the charge density operator $\hat{\rho}(x)$ and the charge current density operator $\hat{\vec{j}}(x)$.

We begin with the Dirac field operator $\hat{\psi}(x)$.
Since the ``rigged" part of the Lagrangian $\hat{L}_a$ does not explicitly depend on  $\hat{\psi}(x)$,
the equation of motion of $\hat{\psi}(x)$ is same as one in the ordinary QED. That is
\begin{eqnarray}
i\hbar \gamma^\mu \hat{D}_{e \mu}(x) \hat{\psi}(x) = m_e c \hat{\psi}(x),
\end{eqnarray}
if written covariantly, and it can be also expressed in a form
\begin{eqnarray}
i\hbar \frac{\partial \hat{\psi}(x)}{\partial t} 
= 
\left\{
 (Z_e e) \hat{A}_0(x) 
   + \vec{\alpha} \cdot \left( -i\hbar c \vec{\nabla} - (Z_e e)  \hat{\vec{A}}(x)  \right)  + m_e c^2 \beta 
   \right\} \hat{\psi}(x),  \label{eq:Dirac} 
\end{eqnarray}
which has the same form as a Hamiltonian form of the Dirac equation. 
In our notation of the gamma matrices, $\beta = \gamma^0$ and $\vec{\alpha} = \gamma^0 \vec{\gamma}$. 

As for the Schr\"{o}dinger field operator $\hat{\chi}_a(x)$, we obtain the equation of motion in the same form 
as the Schr\"{o}dinger equation of the non-relativistic quantum mechanics as 
\begin{eqnarray}
i\hbar \frac{\partial}{\partial t} \hat{\chi}_a(x)
=
-\frac{\hbar^2}{2 m_a} \hat{\vec{D}}^2_a(x) \hat{\chi}_a(x)
+ Z_a e \hat{A}_0(x) \hat{\chi}_a(x). \label{eq:Schrodinger}
\end{eqnarray}
This can also be written as
\begin{eqnarray}
i\hbar \frac{\partial}{\partial t} \hat{\chi}_a(x)
&=&
-\frac{\hbar^2}{2 m_a} 
\left\{
\vec{\nabla}^2 - 2 i\frac{Z_a e}{\hbar c} \hat{\vec{A}}(x) \cdot \vec{\nabla} - \left(\frac{Z_a e}{\hbar c}\right)^2 \hat{\vec{A}}(x) \cdot \hat{\vec{A}}(x)
\right\}
 \hat{\chi}_a(x) \nonumber \\
& &+ Z_a e \hat{A}_0(x) \hat{\chi}_a(x), \label{eq:Schrodinger2}
\end{eqnarray}
where we use the Coulomb gauge condition $\vec{\nabla} \cdot \hat{\vec{A}}(x) = 0$.

Finally, the equations of motion for the photon field $\hat{A}_\mu(x)$ have the same form as the inhomogeneous Maxwell equations
of the classical electrodynamics. In the Coulomb gauge, we have
\begin{eqnarray}
& &- \nabla^2 \hat{A}_0(x)  = 4\pi \hat{\rho}(x), \label{eq:Maxwell1} \\
& &\frac{1}{c} \frac{\partial}{\partial t} \vec{\nabla} \hat{A}_0(x) + \left( \frac{1}{c^2} \frac{\partial^2}{\partial t^2} - \nabla^2 \right) \hat{\vec{A}}(x)
=
\frac{4\pi}{c} \hat{\vec{j}}(x),  \label{eq:Maxwell2} 
\end{eqnarray}
where $\hat{\rho}(x)$ is the charge density operator, $\hat{\vec{j}}(x)$ is the charge current density operator.
Note that in the case of Rigged QED, $\hat{\rho}(x)$ and $\hat{\vec{j}}(x)$ include the contribution from 
atomic nuclei in addition to that of electrons. 
These ``rigged" charge and current are described below. 

The charge density operator $\hat{\rho}(x)$ is the sum of electron charge density operator $\hat{\rho}_e(x)$
and atomic nuclear charge density operator $\hat{\rho}_a(x)$:
\begin{eqnarray}
\hat{\rho}(x) = \hat{\rho}_e(x) + \sum_{a=1}^{N_n} \hat{\rho}_a(x), 
\end{eqnarray}
where
\begin{eqnarray}
\hat{\rho}_e(x) &=& Z_e e \hat{\bar{\psi}}(x) \gamma^0 \hat{\psi}(x), \label{eq:rhoe} \\
\hat{\rho}_a(x) &=& Z_a e \hat{\chi}_a^\dagger(x) \hat{\chi}_a(x).\label{eq:rhoa}
\end{eqnarray}

Similarly, the charge current density operator $\hat{\vec{j}}(x)$ is the sum of electron charge current density operator 
$\hat{\vec{j}}_e(x)$ and atomic nuclear charge current density operator $\hat{\vec{j}}_a(x)$:
\begin{eqnarray}
\hat{\vec{j}}(x) = \hat{\vec{j}}_e(x) + \sum_{a=1}^{N_n} \hat{\vec{j}}_a(x),
\end{eqnarray}
where
\begin{eqnarray}
\hat{\vec{j}}_e(x) &=& Z_e e c\, \hat{\bar{\psi}}(x) \vec{\gamma} \hat{\psi}(x), \label{eq:je} \\
\hat{\vec{j}}_a(x) &=& \frac{ Z_a e}{2m_a} \left( 
i\hbar \hat{\chi}^\dagger_a(x) \hat{\vec{D}}_a(x) \hat{\chi}_a(x) 
- i\hbar \left( \hat{\vec{D}}_a(x) \hat{\chi}_a(x) \right)^\dagger \cdot \hat{\chi}_a(x)
\right). \label{eq:ja}
\end{eqnarray}

Here, in passing, some notes are in order. First, the equations of continuity hold for each species, namely, 
\begin{eqnarray}
\frac{\partial}{\partial t} \hat{\rho}_\alpha(x) + {\rm div} \hat{\vec{j}}_\alpha(x) = 0, 
\end{eqnarray}
where $\alpha=e$ or $a$.
Second, $\hat{\rho}_\alpha(x)$ ($\alpha=e$ or $a$) is connected to
position probability density operator $\hat{N}_\alpha(x)$ as $\hat{\rho}_\alpha(x) = Z_\alpha e \hat{N}_\alpha(x)$ and
$\hat{\vec{j}}_\alpha(x)$ is connected to
velocity density operator $\hat{\vec{v}}_\alpha(x)$ as $\hat{\vec{j}}_\alpha(x) = Z_\alpha e \hat{\vec{v}}_\alpha(x)$.

In summary, the rigged charge and the rigged current are
\begin{eqnarray}
\hat{\rho}(x) 
&=& 
 Z_e e \hat{\bar{\psi}}(x) \gamma^0 \hat{\psi}(x) + \sum_a^{N_n} Z_a e \hat{\chi}_a^\dagger(x) \hat{\chi}_a(x), \label{eq:rhotot} \\
\hat{\vec{j}}(x)
 &=& 
  Z_e e c\, \hat{\bar{\psi}}(x) \vec{\gamma} \hat{\psi}(x) \nonumber \\
& &
 + \sum_a^{N_n} \frac{ Z_a e}{2m_a} \left( 
i\hbar \hat{\chi}^\dagger_a(x) \hat{\vec{D}}_a(x) \hat{\chi}_a(x) 
- i\hbar \left( \hat{\vec{D}}_a(x) \hat{\chi}_a(x) \right)^\dagger \cdot \hat{\chi}_a(x)
\right). \label{eq:jtot}
\end{eqnarray}

Now, we have a closed set of time evolution equations for field operators $\hat{\psi}(x)$, $\hat{\chi}_a(x)$ and $\hat{A}_\mu(x)$:
Eqs.~\eqref{eq:Dirac}, \eqref{eq:Schrodinger2}, \eqref{eq:Maxwell1}, \eqref{eq:Maxwell2}, \eqref{eq:rhotot} and \eqref{eq:jtot}.
To solve them, we rewrite them using annihilation and creation operators as is done in ordinary QED
(or other quantum field theories). 
In the next section, we describe how we expand the field operators by the annihilation and creation operators
and derive time evolution equations for them.

\section{Expansion of field operators by annihilation and creation operators}  \label{sec:expansion}

In this section, we introduce annihilation and creation operators for each field operator and 
describe our expansion method.
As is explained in the following, definitions for annihilation and creation operators are different from those used in the conventional QED treatment.
This reflects our aim to treat dynamics of electrons, photons and atomic nuclei in atomic and molecular systems, which are bound states,
in a non-perturbative manner.


\subsection{Electron field}   \label{sec:expansion_e}

In QED, the electron is expressed by the Dirac field operator $\hat{\psi}(x)$ and it is usually expanded by plane waves, which are solutions for 
the free Dirac equation. 
As is well-established, this works extremely fine for the mostly considered QED processes which are represented by scattering cross sections. 
This is because those processes are described by a perturbation series to non-interacting system with the interaction mediated by photons treated as
the perturbation.

In our case, however, since molecular systems are highly non-perturbative and electrons are bounded, the same method is not likely to work.
In addition, what we want to know is not the cross section, which just describes the transition between infinite past (``in-state") and infinite future (``out-state"),
but time step-by-step evolution of the systems.
To this end, we propose an alternative way to expand  $\hat{\psi}(x)$ as
\begin{eqnarray}
\hat{\psi}(ct, \vec{r}) = \sum_{n=1}^{N_D} \left[ \hat{e}_n(t) \psi_n^{(+)}(\vec{r}) + \hat{f}^\dagger_n(t) \psi_n^{(-)}(\vec{r}) \right],  \label{eq:psi_expand} \\
\hat{\psi}^\dagger(ct, \vec{r}) = \sum_{n=1}^{N_D} \left[ \hat{e}_n^\dagger(t) \psi_n^{\dagger (+)}(\vec{r}) + \hat{f}_n(t) \psi_n^{\dagger (-)}(\vec{r}) \right], 
\end{eqnarray}
where $\hat{e}_n(t)$/$\hat{e}^\dagger_n(t)$ is the electron annihilation/creation operator, $\hat{f}_n(t)$/$\hat{f}^\dagger_n(t)$ is the positron annihilation/creation operator 
and $\psi_n^{(+)}(\vec{r})$ ($\psi_n^{(-)}(\vec{r})$) are the four-component wave functions of the electron (the positron),
which are solutions of the time-independent Dirac equation for a particle in a nucleus field.
Concretely, $\psi_n^{(\pm)}(\vec{r})$ can be considered as the $n$-th molecular orbitals which are obtained by solving the four-component Dirac Hartree-Fock equation 
for the system and $N_D$ is the number of the basis set.

We note that this expansion of the field operator is similar to the Furry representation (picture) 
of ordinary QED \cite{Furry1951, LLRQT}
in a sense that it uses the solutions of the Dirac equation in an external field. 
However, the annihilation/creation operators in the Furry representation do not depend on time (so it can be 
considered as a variant of the interaction picture) whereas those in our expansion carry all the time-dependence of the field operator.
In other words, we adopt to work in Heisenberg picture.
Since we wish to solve the dynamics of the system in a non-perturbative manner,
the use of interaction picture does not make our problem easier.

Before proceeding further, to make equations below less cluttered, we introduce our notation for the annihilation/creation operators and 
 the electron/positron wavefunctions. We define
\begin{eqnarray}
\hat{e}_{n^+} &\equiv& \hat{e}_n , \\
\hat{e}_{n^-} &\equiv& \hat{f}^\dagger_n ,
\end{eqnarray}
and
\begin{eqnarray}
\psi_{n^+}(\vec{r}) &\equiv& \psi^{(+)}_n(\vec{r}) , \\
\psi_{n^-}(\vec{r}) &\equiv& \psi^{(-)}_n(\vec{r}).
\end{eqnarray}
Note that the positron creation operator does not have a dagger as superscript in our notation.
Then the field operator expansion eq.~\eqref{eq:psi_expand} can be written as 
\begin{eqnarray}
\hat{\psi}(ct, \vec{r}) = \sum_{n=1}^{N_D}  \sum_{a=\pm}  \hat{e}_{n^a}(t) \psi_{n^a}(\vec{r}), \label{eq:psi_expand2}
\end{eqnarray}
and orthonormality of the wavefunctions can be expressed as
\begin{eqnarray}
\int d^3\vec{r}\, \psi_{n^a}^\dagger(\vec{r})\psi_{m^b}(\vec{r}) = \delta_{nm} \delta_{ab}.  \label{eq:psi_orthonormality}
\end{eqnarray}
The anti-commutation relation can be written by 
\begin{eqnarray}
\left\{ \hat{e}_{n^a}, \hat{e}_{m^b}^\dagger \right\} = \delta_{nm} \delta_{ab},
\end{eqnarray}
and anti-commutators of other combinations are zero.
Also, for later use, we define the electron excitation operator 
\begin{eqnarray}
\hat{\cal E}_{n^a m^b} \equiv \hat{e}^\dagger_{n^a} \hat{e}_{m^b}.  \label{eq:def_calE}
\end{eqnarray}

\subsection{Nucleus field}  

We expand the Schr\"{o}dinger field operator $\hat{\chi}_a(x)$ for the atomic nucleus $a$ by the 
nucleus annihilation operator $\hat{c}_{a i}$ and creation operator $\hat{c}^\dagger_{a i}$ as
\begin{eqnarray}
\hat{\chi}_a(ct, \vec{r}) = \sum_{i=1}^{N_S} \hat{c}_{ai}(t) \chi_{ai}(\vec{r}), \label{eq:chi_expand} \\
\hat{\chi}^\dagger_a(ct, \vec{r}) = \sum_{i=1}^{N_S} \hat{c}^\dagger_{ai}(t) \chi^*_{ai}(\vec{r}),
\end{eqnarray}
where $\chi_{ai}(\vec{r})$ is a set of orthonormal functions 
\begin{eqnarray}
\int d^3\vec{r}\, \chi_{ai}^*(\vec{r}) \chi_{aj}(\vec{r}) = \delta_{ij}.  \label{eq:chi_orthonormality}
\end{eqnarray}
The commutation relation for $\hat{c}_{a i}$ and $\hat{c}^\dagger_{a i}$ is $\left[\hat{c}_{a i}, \hat{c}^\dagger_{a j} \right] = \delta_{ij}$
and the anti-commutation relation is $\left\{ \hat{c}_{a i}, \hat{c}^\dagger_{a j} \right\} = \delta_{ij}$.
For the other combinations, (anti-)commutators are zero.
Which of them is imposed depends on the $a$'s nuclear spin. 
If $a$ has (half-)integer spin, the (anti-)commutation relation is imposed.

As is done for the electron case, we define the nucleus excitation operator 
\begin{eqnarray}
\hat{\cal C}_{a ij} \equiv \hat{c}^\dagger_{a i} \hat{c}_{a j}, \label{eq:def_calC}
\end{eqnarray}
for the later convenience. 

\subsection{Photon field}  \label{sec:expansion_photon}

In QED, the photon is expressed by a vector field operator $\hat{A}_\mu(x)$ with $U(1)$ gauge symmetry. 
In the standard QED treatment, the quantization is performed to free electromagnetic field and the interaction with the electron is
taken into account as a perturbation.
Although this is not the way we try to solve the dynamics of the system as mentioned previously,
since we are going to use the expression of the quantized free field as a part of  $\hat{A}_\mu(x)$,
we begin by showing its expression. 
We call this $\hat{A}_{{\rm rad}, \mu}(x)$ to distinguish from the total $\hat{A}_\mu(x)$.
In the Coulomb gauge, as is commonly done when quantizing the electromagnetic field (e.g. Ref.~\cite{Peskin, Weinberg, Loudon}), we have
\begin{eqnarray}
\hat{A}^0_{\rm rad}(ct, \vec{r}) &=& 0, \label{eq:Arad0} \\
\hat{A}^k_{\rm rad}(ct, \vec{r})
 &=& \frac{\sqrt{4\pi \hbar^2 c}}{\sqrt{(2\pi\hbar)^3}} \sum_{\sigma=\pm1} \int \frac{d^3 \vec{p}}{\sqrt{2p^0}}
\bigg[ \hat{a}(\vec{p},\sigma) e^k(\vec{p},\sigma) e^{-i c p^0 t/\hbar} e^{i \vec{p}\cdot \vec{r} /\hbar} \nonumber \\
& &+ \hat{a}^\dagger(\vec{p},\sigma) e^{* k}(\vec{p},\sigma) e^{i c p^0 t/\hbar} e^{-i \vec{p}\cdot \vec{r} /\hbar} \bigg] ,
\label{eq:Aradvec}
\end{eqnarray}
where $\hat{a}(\vec{p},\sigma)$ is the annihilation operator of the photon with momentum $\vec{p}$ and helicity $\sigma$ and
$\vec{e}$ is the polarization vector. The photon annihilation/creation operators satisfy the commutation relation 
\begin{eqnarray}
[ \hat{a}(\vec{p},\sigma), \hat{a}^\dagger(\vec{p}\,', \sigma') ] = \delta(\vec{p}-\vec{p}\,') \delta_{\sigma \sigma'},
\end{eqnarray}
and commutators of the other combinations are zero. 
As for the polarization vector, when we take the photon momentum in the polar coordinate as 
\begin{eqnarray}
\vec{p}= 
\begin{pmatrix}
p^0 \sin \theta \cos \phi \\
p^0 \sin \theta \sin \phi \\
p^0 \cos \theta
\end{pmatrix},
\end{eqnarray}
($0 \le \theta \le \pi$, $0 \le \phi < 2\pi$, $|\vec{p}| = p^0$), we have
\begin{eqnarray}
\vec{e}(\vec{p},\sigma) = \frac{1}{\sqrt{2}} 
\begin{pmatrix}
\cos \phi \cos \theta \mp i \sin \phi \\
\sin \phi \cos \theta \pm i \cos \phi \\
-\sin \theta
\end{pmatrix},
\end{eqnarray}
where the double sign corresponds to $\sigma = \pm 1$.
We note that $\hat{A}^0_{\rm rad}(ct, \vec{r})$ and $\hat{\vec{A}}_{\rm rad}(ct, \vec{r})$ by construction 
satisfy Maxwell equations \eqref{eq:Maxwell1} and \eqref{eq:Maxwell2} with the right-hand-sides being zero ($i.e.$ equations for free fields).

To solve the dynamics of the system non-perturbatively, we take advantage of the fact that the formal solutions of the inhomogeneous 
Maxwell equations \eqref{eq:Maxwell1} and \eqref{eq:Maxwell2} are known from the classical electrodynamics \cite{LLCTF}. 
We shall employ the strategy to express $\hat{A}_\mu(x)$ by such solutions \cite{Tachibana2003}. 

As for the first inhomogeneous Maxwell equation \eqref{eq:Maxwell1}, since it only contains the scalar potential $\hat{A}^0(ct, \vec{r})$,
its solution is readily written by the sum of the solution of Eq.~\eqref{eq:Maxwell1} with the right-hand-side being zero and the particular solution
of Eq.~\eqref{eq:Maxwell1}. The former is $\hat{A}^0_{\rm rad}(ct, \vec{r})$ which is zero (Eq.~\eqref{eq:Arad0}) and the latter is known as
the solution of the Poisson equation. Thus,
\begin{eqnarray}
\hat{A}_0(ct, \vec{r}) = \int d^3\vec{s}\, \frac{\hat{\rho}(ct, \vec{s})}{|\vec{r}-\vec{s}|}. \label{eq:A0}
\end{eqnarray}

As for the second inhomogeneous Maxwell equation \eqref{eq:Maxwell2}, we can make it simpler by decomposing the current $\hat{\vec{j}}(x)$
into the transversal component $\hat{\vec{j}}_T(x)$ and the longitudinal component $\hat{\vec{j}}_L(x)$ as
\begin{eqnarray}
\hat{\vec{j}}(x) &=& \hat{\vec{j}}_T(x) + \hat{\vec{j}}_L(x),
\end{eqnarray}
where $\vec{\nabla} \cdot  \hat{\vec{j}}_T(x) = 0$ and $\vec{\nabla} \times  \hat{\vec{j}}_L(x) = 0$.
This transforms Eq.~\eqref{eq:Maxwell2} into two equations each involving only $\hat{A}^0(ct, \vec{r})$ or $\hat{\vec{A}}(ct, \vec{r})$ 
\begin{eqnarray}
\left( \frac{1}{c^2} \frac{\partial^2}{\partial t^2} - \nabla^2 \right) \hat{\vec{A}}(x)
=
\frac{4\pi}{c} \hat{\vec{j}}_T(x), \label{eq:Maxwell2a} \\
\frac{\partial}{\partial t} \vec{\nabla} \hat{A}_0(x) = 4\pi \hat{\vec{j}}_L(x).\label{eq:Maxwell2b}
\end{eqnarray}
Since we have already found $\hat{A}_0(x)$ as Eq.~\eqref{eq:A0}, Eq.~\eqref{eq:Maxwell2b} tells us $\hat{\vec{j}}_L(x)$,
which in turn gives the right-hand-side of Eq.~\eqref{eq:Maxwell2a} by  $\hat{\vec{j}}(\vec{r}) - \hat{\vec{j}}_L(\vec{r})$.
Then we can obtain the solution of Eq.~\eqref{eq:Maxwell2a} by 
the sum of the solution of Eq.~\eqref{eq:Maxwell2a} with the right-hand-side being zero and the particular solution
of Eq.~\eqref{eq:Maxwell2a}. 
The former is $\hat{\vec{A}}_{\rm rad}(ct, \vec{r})$ (Eq.~\eqref{eq:Aradvec}) which is written by the annihilation/creation operators and
the latter is known from the classical electromagnetism, which we shall call $\hat{\vec{A}}_A(ct, \vec{r})$. 
Written explicitly, 
\begin{eqnarray}
\hat{\vec{A}}_A(ct, \vec{r}) = \frac{1}{c} \int d^3\vec{s}\, \frac{\hat{\vec{j}}_T(c u, \vec{s})}{|\vec{r}-\vec{s}|},  \label{eq:A_A}
\end{eqnarray}
where
\begin{eqnarray}
u = t - \frac{|\vec{r}-\vec{s}|}{c}. \label{eq:retardation}
\end{eqnarray}
In summary, the vector potential $\hat{\vec{A}}(ct, \vec{r})$ is  
\begin{eqnarray}
\hat{\vec{A}}(ct, \vec{r}) = \hat{\vec{A}}_{\rm rad}(ct, \vec{r}) + \hat{\vec{A}}_A(ct, \vec{r}),
\end{eqnarray}
which are given by Eqs.~\eqref{eq:Aradvec}, \eqref{eq:A_A} and \eqref{eq:retardation}.

\subsection{Charge density and charge current density}  \label{sec:charge}

In this subsection we express the charge density operator $\hat{\rho}(x)$ (Eq.~\eqref{eq:rhotot}) and
 the charge current density operator $\hat{\vec{j}}(x)$  (Eq.~\eqref{eq:jtot}) in terms of 
annihilation and creation operators (or excitation operators) which are defined in previous subsections. 

The expression for the electronic charge density operator can be derived by substituting Eq.~\eqref{eq:psi_expand2} into 
Eq.~\eqref{eq:rhoe}, and the atomic nucleus charge density operator by Eq.~\eqref{eq:chi_expand} into Eq.~\eqref{eq:rhoa}.
They give
\begin{eqnarray}
\hat{\rho}_e(x) &=& \sum_{p,q=1}^{N_D} \sum_{c,d=\pm} \rho_{p^c q^d}(\vec{r}) \hat{\cal E}_{p^c q^d}, \label{eq:rhoe2} \\
\hat{\rho}_a(x) &=&  \sum_{i,j=1}^{N_S}  \rho_{a ij}(\vec{r}) \hat{\cal C}_{a ij},\label{eq:rhoa2}
\end{eqnarray}
where
\begin{eqnarray}
\rho_{p^c q^d}(\vec{r}) &\equiv& (Z_e e) \psi_{p^c}^\dagger(\vec{r})  \psi_{q^d}(\vec{r}),  \label{eq:func_rhoe} \\
\rho_{a ij}(\vec{r}) &\equiv& (Z_a e) \chi^*_{ai}(\vec{r}) \chi_{aj}(\vec{r}). \label{eq:func_rhoa}
\end{eqnarray}

The expression for the charge current density operator can be derived in a similar manner. 
From Eq.~\eqref{eq:je}, the electronic charge current density operator is 
\begin{eqnarray}
\hat{j}_e^k(x) &=& \sum_{p,q=1}^{N_D} \sum_{c,d=\pm}  j^k_{p^c q^d}(\vec{r}) \hat{\cal E}_{p^c q^d},
\end{eqnarray}
where
\begin{eqnarray}
j^k_{p^c q^d}(\vec{r}) &\equiv& Z_e\, e\,c \left[ \psi_{p^c}^\dagger(\vec{r}) \gamma^0 \gamma^k \psi_{q^d}(\vec{r})  \right].  \label{eq:func_jk}
\end{eqnarray}
From Eq.~\eqref{eq:ja}, the atomic nucleus charge current density operator is 
\begin{eqnarray}
\hat{j}_a^k(x)  
&=&
\sum_{i,j=1}^{N_S} \left\{ j_{aij}^k(\vec{r}) \hat{\cal C}_{aij}
-  \frac{(Z_a e)}{m_a c} \rho_{aij}(\vec{r}) \left( \hat{A}_{\rm rad}^k \hat{\cal C}_{aij} + \hat{c}_{ai}^\dagger \hat{A}_A^k \hat{c}_{aj} \right) \right\}, 
\end{eqnarray}
where
\begin{eqnarray}
j_{a ij}^k(\vec{r}) \equiv - (Z_a e) \frac{i\hbar }{2m_a} \left\{ \chi_{ai}^*(\vec{r}) \nabla^k \chi_{aj}(\vec{r}) - \left( \nabla^k \chi_{ai}^*(\vec{r}) \right) \chi_{aj}(\vec{r}) \right\}. \label{eq:func_jak}
\end{eqnarray}
Note that in our notation the spatial components of the covariant derivative are $\vec{D}(x) = -\vec{\nabla}  + i\frac{q}{\hbar c} \vec{A}(x)$. 
Also, we have used the fact that $\hat{c}_{aj}$ commutes with $\hat{A}_{\rm rad}^k$ but not with $\hat{A}_A^k$.

Therefore, the total charge density operator $\hat{\rho}(x)$  and
 the charge current density operator $\hat{\vec{j}}(x)$ are
 \begin{eqnarray}
\hat{\rho}(x) &=& 
\sum_{p,q=1}^{N_D} \sum_{c,d=\pm} \rho_{p^c q^d}(\vec{r}) \hat{\cal E}_{p^c q^d} 
+ \sum_{a=1}^{N_n} \sum_{i,j=1}^{N_S}  \rho_{a ij}(\vec{r}) \hat{\cal C}_{a ij},  \label{eq:rhotot2} \\
\hat{j}^k(x) &=&
\sum_{p,q=1}^{N_D} \sum_{c,d=\pm}  j^k_{p^c q^d}(\vec{r}) \hat{\cal E}_{p^c q^d} \nonumber \\
& &+ \sum_{a=1}^{N_n} \sum_{i,j=1}^{N_S}  \left\{ j_{aij}^k(\vec{r}) \hat{\cal C}_{aij}
-  \frac{(Z_a e)}{m_a c} \rho_{aij}(\vec{r}) \left( \hat{A}_{\rm rad}^k \hat{\cal C}_{aij} + \hat{c}_{ai}^\dagger \hat{A}_A^k \hat{c}_{aj} \right) \right\}. \label{eq:jtot2}
\end{eqnarray}

Now, we can rewrite the scalar potential $\hat{A}^0(ct, \vec{r})$ using the excitation operators. 
Substituting Eq.~\eqref{eq:rhotot2} into Eq.~\eqref{eq:A0} gives
\begin{eqnarray}
\hat{A}_0(ct, \vec{r})
 =
  \sum_{p,q=1}^{N_D} \sum_{c,d=\pm} V_{p^c q^d}(\vec{r}) \hat{\cal E}_{p^c q^d}
 + \sum_{a=1}^{N_n}\sum_{i,j=1}^{N_S} V_{a ij}(\vec{r}) \hat{\cal C}_{a ij},   \label{eq:A0_2}
\end{eqnarray}
where we define integrals
\begin{eqnarray}
V_{p^c q^d}(\vec{R})
&\equiv&
  \int d^3\vec{s}\, \frac{\rho_{p^c q^d}(\vec{s}) }{|\vec{s}-\vec{R}|},
\label{eq:intVe} \\
V_{a ij}(\vec{R}) 
&\equiv& 
\int d^3\vec{s}\, \frac{\rho_{a ij}(\vec{s})}{|\vec{s}-\vec{R}|}.
\label{eq:intVa}
\end{eqnarray}
Then, we can express $\hat{\vec{j}}_L(x)$ using the excitation operators via Eq.~\eqref{eq:Maxwell2b}.
Substituting Eq.~\eqref{eq:A0_2} into Eq.~\eqref{eq:Maxwell2b} gives
\begin{eqnarray}
\hat{j}^k_L(x) 
&=& 
-\sum_{p,q=1}^{N_D} \sum_{c,d=\pm}  E^k_{p^c q^d}(\vec{r}) \frac{d \hat{\cal E}_{p^c q^d}}{dt} 
-\sum_{a=1}^{N_n}\sum_{i,j=1}^{N_S}  E^k_{a ij}(\vec{r}) \frac{d \hat{\cal C}_{a ij}}{dt},   \label{eq:jL}
\end{eqnarray}
where we define integrals
\begin{eqnarray}
E^{k}_{p^c q^d}(\vec{R}) 
&\equiv& 
-\frac{1}{4\pi} \frac{\partial}{\partial R^k}V_{p^c q^d}(\vec{R}) 
=
-\frac{Z_e e}{4\pi} \int d^3\vec{s}\, \psi_{p^c}^\dagger(\vec{s}) \psi_{q^d}(\vec{s})  \frac{(\vec{s}-\vec{R})^k}{|\vec{s}-\vec{R}|^3},   \label{eq:intEe}  \\
E^{k}_{a ij}(\vec{R}) 
&\equiv& 
-\frac{1}{4\pi} \frac{\partial}{\partial R^k}V_{a ij}(\vec{R}) 
= 
-\frac{Z_a e}{4\pi} \int d^3\vec{s}\, \chi^*_{a i}(\vec{s}) \chi_{a j}(\vec{s})  \frac{(\vec{s}-\vec{R})^k}{|\vec{s}-\vec{R}|^3}.   
\label{eq:intEa} 
\end{eqnarray}

Finally, we can obtain $\hat{\vec{j}}_T(x)$ from Eqs.~\eqref{eq:jtot2} and \eqref{eq:jL}, and in turn $\hat{\vec{A}}_A(ct, \vec{r})$ using Eq.~\eqref{eq:A_A}.
However, the existence of retardation (Eq.~\eqref{eq:retardation}) in the right-hand-side of Eq.~\eqref{eq:A_A} prevent us from 
making the expression of $\hat{\vec{A}}_A(ct, \vec{r})$ simpler. 
In other words, $\hat{\vec{A}}_A(ct, \vec{r})$ is determined by $\hat{\vec{j}}_T$ at earlier times than $t$,
whose expression contains $\hat{\vec{A}}_A$ at these times. 
Thus, $\hat{\vec{A}}_A(ct, \vec{r})$ is computed step-by-step in time using all the information of itself at earlier times. 

\section{Time evolution of annihilation and creation operators}  \label{sec:evolution}

In this section, we derive time evolution equations for annihilation operators of the electron $\hat{e}_{p^c}$ and the nucleus $\hat{c}_{a i}$.
Those for the creation operators can be obtained by taking Hermite conjugate. 
We note that in our formalism, the photon annihilation operator  $\hat{a}(\vec{p},\sigma)$ does not depend on time as is in the standard QED treatment.
We also discuss the time evolution of the excitation operators and physical quantities. 

\subsection{Electron annihilation operator}  

We first derive the time derivative of the electron annihilation operator $\hat{e}_{p^c}$. 
This is obtained by substituting the expansion \eqref{eq:psi_expand2} into \eqref{eq:Dirac},
multiplying by $\psi^\dagger_{p^c}(\vec{r})$ and integrating over $\vec{r}$. 
Then, the orthonormality of the wavefunctions (Eq.~\eqref{eq:psi_orthonormality}) yields
\begin{eqnarray}
i\hbar \frac{\partial \hat{e}_{p^c}}{\partial t} 
&=&
\sum_{q=1}^{N_D}  \sum_{d=\pm} \left( \hat{I}_{1 p^c q^d} + \hat{I}_{2 p^c q^d} + \hat{I}_{3 p^c q^d} + \hat{I}_{4 p^c q^d}\right)  \hat{e}_{q^d} , \label{eq:dedt}
\end{eqnarray}
where
\begin{eqnarray}
\hat{I}_{1 p^c q^d}
&=&
\int d^3\vec{r}\,  \psi^\dagger_{p^c}(\vec{r}) \vec{\alpha} \cdot \left( -i\hbar c \vec{\nabla} \right) \psi_{q^d}(\vec{r}),
\\
\hat{I}_{2 p^c q^d}
&=&
\int d^3\vec{r}\,  \psi^\dagger_{p^c}(\vec{r}) \vec{\alpha} \cdot \left(  - (Z_e e)  \hat{\vec{A}}(x)  \right) \psi_{q^d}(\vec{r}),
\\
\hat{I}_{3 p^c q^d}
&=&
\int d^3\vec{r}\,  \psi^\dagger_{p^c}(\vec{r})( m_e c^2 )\beta \psi_{q^d}(\vec{r}),
\\
\hat{I}_{4 p^c q^d}
&=&
\int d^3\vec{r}\,  \psi^\dagger_{p^c}(\vec{r}) (Z_e e) \hat{A}_0(x) \psi_{q^d}(\vec{r}).
\end{eqnarray}

$\hat{I}_{1 p^c q^d}$ and $\hat{I}_{3 p^c q^d}$ are contributions from kinetic energy and mass energy respectively.
We define the electron kinetic energy integral
\begin{eqnarray}
T_{p^c q^d} \equiv  -i\hbar c \int d^3\vec{r}\,\psi_{p^c}^\dagger(\vec{r}) \gamma^0  \vec{\gamma} \cdot \vec{\nabla} \psi_{q^d}(\vec{r}),
\label{eq:intT}
\end{eqnarray}
and the electron mass energy integral
\begin{eqnarray}
M_{p^c q^d} 
\equiv m_e c^2 \int d^3\vec{r}\,\psi_{p^c}^\dagger(\vec{r}) \gamma^0  \psi_{q^d}(\vec{r}).
\label{eq:intM}
\end{eqnarray}
Then, we have $\hat{I}_{1 p^c q^d} = T_{p^c q^d}$ and $\hat{I}_{3 p^c q^d}=M_{p^c q^d}$.

$\hat{I}_{4 p^c q^d}$ can be rewritten using Eq.~\eqref{eq:A0_2} as
\begin{eqnarray}
\hat{I}_{4 p^c q^d} 
&=& 
\int d^3\vec{r}\,  \rho_{p^c q^d}(\vec{r}) \hat{A}_0(x) \\
&=&
   \sum_{r,s=1}^{N_D} \sum_{e,f=\pm}   (p^c q^d | r^e s^f ) \hat{\cal E}_{r^e s^f}
 + \sum_{a=1}^{N_n}\sum_{i,j=1}^{N_S} (p^c q^d | i_a j_a)  \hat{\cal C}_{aij},  
 \label{eq:I4pq}
\end{eqnarray}
where we define two types of four-center integral as
\begin{eqnarray}
(p^c q^d | r^e s^f) 
&\equiv&
(Z_e e)^2 \int d^3\vec{r}\, d^3\vec{s}\, \psi_{p^c}^\dagger(\vec{r}) \psi_{q^d}(\vec{r}) 
\frac{1}{|\vec{r}-\vec{s}|} \psi_{r^e}^\dagger(\vec{s}) \psi_{s^f}(\vec{s}), \label{eq:int4c1} \\
(p^c q^d | i_a j_a ) 
&\equiv&
  (Z_e  e) (Z_a e)   \int d^3\vec{r}\,  d^3\vec{s}\,  \psi_{p^c}^\dagger(\vec{r}) \psi_{q^d}(\vec{r}) \frac{1}{|\vec{r}-\vec{s}|} \chi^*_{ai}(\vec{s}) \chi_{aj}(\vec{s}). \label{eq:int4c2}
\end{eqnarray}

We can write $\hat{I}_{2 p^c q^d}$ using Eq.~\eqref{eq:func_jk} as
\begin{eqnarray}
\hat{I}_{2 p^c q^d} 
&=& 
 -\frac{1}{c} \int d^3\vec{r}\, \vec{j}_{p^c q^d}(\vec{r}) \cdot \left(\hat{\vec{A}}_{\rm rad}(x)+\hat{\vec{A}}_A(x) \right).  \label{eq:I2e}
 \end{eqnarray}
Here, the term including $\hat{\vec{A}}_{\rm rad}(x)$ can be written, by substituting Eq.~\eqref{eq:Aradvec}, as
\begin{eqnarray}
& & \hspace{-1.5cm} -\frac{1}{c} \int d^3\vec{r}\, \vec{j}_{p^c q^d}(\vec{r}) \cdot \hat{\vec{A}}_{\rm rad}(x) \nonumber \\
&=&
 -\frac{1}{c} \frac{\sqrt{4\pi \hbar^2 c}}{\sqrt{(2\pi\hbar)^3}}
  \sum_{\sigma=\pm1} \int \frac{d^3 \vec{p}}{\sqrt{2p^0}} \times \nonumber \\
& &\left[ 
\vec{F}_{p^c q^d}(\vec{p}) \cdot \vec{e}(\vec{p},\sigma) e^{-i c p^0 t/\hbar} \hat{a}(\vec{p},\sigma)
+\vec{F}_{p^c q^d}(-\vec{p}) \cdot \vec{e^*}(\vec{p},\sigma) e^{i c p^0 t/\hbar}   \hat{a}^\dagger(\vec{p},\sigma)   
\right],   \label{eq:I2e_Arad}
\end{eqnarray} 
where we define the integral which is the Fourier transformation of the function for the electron charge current density (Eq.~\eqref{eq:func_jk}) 
\begin{eqnarray}
F^k_{p^c q^d}(\vec{p}) 
\equiv \int d^3\vec{r}\,  j^k_{p^c q^d}(\vec{r}) e^{i \vec{p}\cdot \vec{r} /\hbar}  .
\label{eq:intF}
\end{eqnarray}
The term including $\hat{\vec{A}}_A(x)$ can be written, by substituting Eq.~\eqref{eq:A_A}, as
\begin{eqnarray}
  -\frac{1}{c} \int d^3\vec{r}\, \vec{j}_{p^c q^d}(\vec{r}) \cdot \hat{\vec{A}}_A(x) 
&=&
 -\frac{1}{c^2} \int d^3\vec{r}\,  d^3\vec{s}\, \frac{\vec{j}_{p^c q^d}(\vec{r}) \cdot \hat{\vec{j}}_T(c u, \vec{s})}{|\vec{r}-\vec{s}|}.
  \label{eq:I2e_AA}
\end{eqnarray}
Further simplification is not possible due to the retardation $u = t - \frac{|\vec{r}-\vec{s}|}{c}$ in $\hat{\vec{j}}_T$.

Putting these all together, we obtain
\begin{eqnarray}
i\hbar \frac{\partial \hat{e}_{p^c}}{\partial t} 
&=&
\sum_{q=1}^{N_D}  \sum_{d=\pm} (T_{p^c q^d}+ M_{p^c q^d} ) \hat{e}_{q^d}
+
\sum_{q,r,s=1}^{N_D}  \sum_{d,e,f=\pm}  (p^c q^d | r^e s^f ) \hat{\cal E}_{r^e s^f} \hat{e}_{q^d} \nonumber \\
&+& \sum_{q=1}^{N_D}  \sum_{d=\pm} \sum_{a=1}^{N_n}\sum_{i,j=1}^{N_S} (p^c q^d | i_a j_a)  \hat{\cal C}_{aij} \hat{e}_{q^d}  \nonumber \\
& -&\frac{1}{c^2}  \sum_{q=1}^{N_D}  \sum_{d=\pm} \int d^3\vec{r}\,  d^3\vec{s}\, \frac{\vec{j}_{p^c q^d}(\vec{r}) \cdot \hat{\vec{j}}_T(c u, \vec{s})}{|\vec{r}-\vec{s}|}    \hat{e}_{q^d} \nonumber \\
&-&\frac{1}{c} \frac{\sqrt{4\pi \hbar^2 c}}{\sqrt{(2\pi\hbar)^3}}
\sum_{q=1}^{N_D}  \sum_{d=\pm}  \sum_{\sigma=\pm1} \int \frac{d^3 \vec{p}}{\sqrt{2p^0}} \times \nonumber \\
& &\hspace{-0.5cm} \left[ 
\vec{F}_{p^c q^d}(\vec{p}) \cdot \vec{e}(\vec{p},\sigma) e^{-i c p^0 t/\hbar} \hat{a}(\vec{p},\sigma)\hat{e}_{q^d}
+\vec{F}_{p^c q^d}(-\vec{p}) \cdot \vec{e^*}(\vec{p},\sigma) e^{i c p^0 t/\hbar}   \hat{a}^\dagger(\vec{p},\sigma)   \hat{e}_{q^d}
\right].
\label{eq:dedt2}
\end{eqnarray}
The Born-Oppenheimer approximation version of this equation is derived in the Appendix~\ref{sec:BO}.

\subsection{Nucleus annihilation operator}  

We next derive the time derivative of the nucleus annihilation operator $\hat{c}_{a i}$. 
Similarly to the electron case, 
this is obtained by substituting the expansion \eqref{eq:chi_expand} into \eqref{eq:Schrodinger2},
multiplying by $\chi^*_{a i}(\vec{r})$ and integrating over $\vec{r}$. 
Then, the orthonormality of the expansion functions (Eq.~\eqref{eq:chi_orthonormality}) yields
\begin{eqnarray}
i\hbar \frac{\partial  \hat{c}_{ai}}{\partial t}
&=&
\sum_{j=1}^{N_S} 
\left(
\hat{I}_{1aij} + \hat{I}_{2aij}+ \hat{I}_{3aij} + \hat{I}_{4aij}
\right)
\hat{c}_{aj},   \label{eq:dcdt}
\end{eqnarray}
where
\begin{eqnarray}
\hat{I}_{1aij}
&=&
-\frac{\hbar^2}{2m_a} \int d^3\vec{r}\, \chi^*_{ai}(\vec{r}) \vec{\nabla}^2 \chi_{aj}(\vec{r}),
\\
\hat{I}_{2aij}
&=&
\frac{i\hbar Z_a e}{m_a c} \int d^3\vec{r}\, \hat{\vec{A}}(x) \cdot  \left(\chi^*_{ai}(\vec{r})  \vec{\nabla} \chi_{aj}(\vec{r}) \right),
\\
\hat{I}_{3aij}
&=&
\frac{(Z_a e)^2}{2 m_a c^2} \int d^3\vec{r}\, \hat{\vec{A}}(x) \cdot \hat{\vec{A}}(x) \chi^*_{ai}(\vec{r}) \chi_{aj}(\vec{r}),
\\
\hat{I}_{4aij}
&=&
(Z_a e) \int d^3\vec{r}\, \hat{A}_0(x) \chi^*_{ai}(\vec{r}) \chi_{aj}(\vec{r}).
\end{eqnarray}

$\hat{I}_{1aij}$ is the contribution from nucleus kinetic energy. Defining the nucleus kinetic energy integral by
\begin{eqnarray}
T_{aij}
&=&
-\frac{\hbar^2}{2m_a} \int d^3\vec{r}\, \chi^*_{ai}(\vec{r}) \vec{\nabla}^2 \chi_{aj}(\vec{r}),
\end{eqnarray}
we have $\hat{I}_{1aij}=T_{aij}$.

$\hat{I}_{4 aij}$  can be rewritten using Eq.~\eqref{eq:A0_2}.
\begin{eqnarray}
\hat{I}_{4aij}
&=&
\int d^3\vec{r}\,  \rho_{aij}(\vec{r}) \hat{A}_0(x) \\
&=&
 \sum_{p,q=1}^{N_D} \sum_{c,d=\pm} (p^c q^d | i_a j_a ) \hat{\cal E}_{p^c q^d} 
 + \sum_{b=1}^{N_n} \sum_{k,l=1}^{N_S} (i_a j_a | k_b l_b)  \hat{\cal C}_{bkl}, \label{eq:I4aij}
\end{eqnarray}
where we use Eq.~\eqref{eq:int4c2} and define another four-center integral 
\begin{eqnarray}
(i_a j_a | k_b l_b) 
&\equiv&
(Z_a e)   (Z_b  e) \int d^3\vec{r}\, d^3\vec{s}\,   \chi^*_{ai}(\vec{r}) \chi_{aj}(\vec{r})\frac{1}{|\vec{r}-\vec{s}|}  \chi^*_{bk}(\vec{s}) \chi_{bl}(\vec{s}).  \label{eq:int4c4}
\end{eqnarray}

$\hat{I}_{2 aij}$  can be written as
\begin{eqnarray}
\hat{I}_{2 aij} 
&=& 
 -\frac{1}{c} \int d^3\vec{r}\, \vec{j}_{aij}(\vec{r}) \cdot \left(\hat{\vec{A}}_{\rm rad}(x)+\hat{\vec{A}}_A(x) \right),
 \end{eqnarray}
where we integrate by parts, use the Coulomb gauge condition and Eq.~\eqref{eq:func_jak}.
We note that this has the same form as Eq.~\eqref{eq:I2e} but  $\vec{j}_{p^c q^d}(\vec{r})$ replaced by $\vec{j}_{aij}(\vec{r})$.
Then the term including $\hat{\vec{A}}_{\rm rad}(x)$ should have the right-hand-side of Eq.~\eqref{eq:I2e_Arad} but 
$\vec{F}_{p^c q^d}(\vec{p})$ (Eq.~\eqref{eq:intF}) replaced by $\vec{F}_{aij}(\vec{p})$ where
\begin{eqnarray}
F^k_{aij}(\vec{p}) 
\equiv \int d^3\vec{r}\,  j^k_{aij}(\vec{r}) e^{i \vec{p}\cdot \vec{r} /\hbar},  
\label{eq:intFa}
\end{eqnarray}
is the Fourier transformation of the function for the nucleus charge current density (Eq.~\eqref{eq:func_jak}). 
Also, the term including $\hat{\vec{A}}_A(x)$ is Eq.~\eqref{eq:I2e_AA} but $\vec{j}_{p^c q^d}(\vec{r})$ replaced by $\vec{j}_{aij}(\vec{r})$.

$\hat{I}_{3 aij}$ can be decomposed into four terms because $\hat{\vec{A}}_{\rm rad}(x)$ and $\hat{\vec{A}}_A(x)$ do not commute.
Namely,
\begin{eqnarray}
\hat{I}_{3 aij} 
=
 \frac{Z_a e}{2 m_a c^2} \int d^3\vec{r}\, 
\left(
\hat{\vec{A}}_{\rm rad} \cdot \hat{\vec{A}}_{\rm rad}
+ \hat{\vec{A}}_{\rm rad} \cdot \hat{\vec{A}}_A
+  \hat{\vec{A}}_A\cdot \hat{\vec{A}}_{\rm rad} 
+ \hat{\vec{A}}_A \cdot \hat{\vec{A}}_A
\right)
\rho_{a ij}(\vec{r}).
 \end{eqnarray}
Since the last three terms involve $\hat{\vec{A}}_A(x)$ which includes the retarded potential, we cannot make their expression simpler. 
The first term can be rewritten using the Fourier transformation of the function for the nucleus charge density (Eq.~\eqref{eq:func_rhoa}),
\begin{eqnarray}
F_{aij}(\vec{p}) 
\equiv \int d^3\vec{r}\,  \rho_{aij}(\vec{r}) e^{i \vec{p}\cdot \vec{r} /\hbar},  
\label{eq:intFrhoa}
\end{eqnarray}
and subsequently defining 
\begin{eqnarray}
G_{aij}(\vec{p},\sigma,\vec{q}, \tau)  \equiv  \vec{e}(\vec{p},\sigma) \cdot \vec{e}(\vec{q},\tau)\, F_{aij}(\vec{p}+\vec{q}),
\end{eqnarray}
as 
\begin{eqnarray}
 \frac{Z_a e}{2 m_a c^2} \int d^3\vec{r}\, 
\left(
\hat{\vec{A}}_{\rm rad} \cdot \hat{\vec{A}}_{\rm rad}
\right)
\rho_{a ij}(\vec{r}) 
&=&
\frac{Z_a e}{4\pi^2 m_a \hbar c} \sum_{\sigma,\tau=\pm 1} 
\int \frac{d^3 \vec{p}}{\sqrt{2p^0}} \frac{d^3 \vec{q}}{\sqrt{2q^0}} \times
\nonumber \\
& & \hspace{-7.5cm} \bigg[ 
G_{aij}(\vec{p},\sigma,\vec{q}, \tau) e^{-ic(p^0+q^0)t/\hbar} \hat{a}(\vec{p},\sigma)\hat{a}(\vec{q},\tau)
+ G_{aij}(\vec{p},\sigma,-\vec{q}, \tau) e^{-ic(p^0-q^0)t/\hbar} \hat{a}(\vec{p},\sigma)\hat{a}^\dagger(\vec{q},\tau)  + \nonumber \\
& & \hspace{-7.5cm} G_{aij}(-\vec{p},\sigma,\vec{q}, \tau) e^{ic(p^0-q^0)t/\hbar} \hat{a}^\dagger(\vec{p},\sigma)\hat{a}(\vec{q},\tau)
+ G_{aij}(-\vec{p},\sigma,-\vec{q}, \tau) e^{ic(p^0+q^0)t/\hbar} \hat{a}^\dagger(\vec{p},\sigma)\hat{a}^\dagger(\vec{q},\tau) 
\bigg],
 \end{eqnarray}
where we used $e^\mu(-\vec{p},\sigma) = e^{*\mu}(\vec{p},\sigma)$.

\subsection{Excitation operators and physical quantities}    \label{sec:evolution_exphys}

As is shown in Eqs.~\eqref{eq:rhoe2} and \eqref{eq:rhoa2}, operators for physical quantities are expressed by the excitation operators 
$\hat{\cal E}_{p^c q^d}$ (Eq.~\eqref{eq:def_calE}) and $\hat{\cal C}_{a ij}$ (Eq.~\eqref{eq:def_calC}). 
(For the field operator expression of other physical quantities of our interests, we refer Refs.~\cite{Tachibana2001,Tachibana2010}.)
Thus, we here show the time derivative of $\hat{\cal E}_{p^c q^d}$ and $\hat{\cal C}_{a ij}$. 

As for $\hat{\cal E}_{p^c q^d}$, Eqs.~\eqref{eq:def_calE} and \eqref{eq:dedt} (and its Hermite conjugate) give
\begin{eqnarray}
i\hbar \frac{d \hat{\cal E}_{p^c q^d}}{dt} 
&=&
\sum_{r=1}^{N_D}  \sum_{e=\pm}
\left( - \hat{e}^\dagger_{r^e} \hat{I}_{r^e p^c} \hat{e}_{q^d}  + \hat{e}^\dagger_{p^c} \hat{I}_{q^d r^e}  \hat{e}_{r^e}  \right),
\label{eq:dcalEdt}
\end{eqnarray}
where we define $\hat{I}_{p^c q^d}  \equiv \hat{I}_{1p^c q^d} + \hat{I}_{2p^c q^d}+ \hat{I}_{3p^c q^d} + \hat{I}_{4p^c q^d}$ and
use $\hat{I}^\dagger_{p^c q^d} = \hat{I}_{q^d p^c}$.
Similarly, Eqs.~\eqref{eq:def_calC} and \eqref{eq:dcdt} (and its Hermite conjugate) give
\begin{eqnarray}
i\hbar \frac{d \hat{\cal C}_{a ij}}{dt} 
&=&
\sum_{k=1}^{N_S} 
\left( - \hat{c}^\dagger_{a k} \hat{I}_{aki} \hat{c}_{a j}  + \hat{c}^\dagger_{a i} \hat{I}_{ajk}  \hat{c}_{a k}   \right)
\label{eq:dcalCdt}
\end{eqnarray}
where we define $\hat{I}_{aij}  \equiv \hat{I}_{1aij} + \hat{I}_{2aij}+ \hat{I}_{3aij} + \hat{I}_{4aij}$ and 
use $\hat{I}_{aij}^\dagger = \hat{I}_{aji}$.

Finally, we describe how we can calculate physical quantities from these equations taking the electronic charge density for example.
The electronic  charge density operator is shown in Eq.~\eqref{eq:rhoe2}. 
To make the operator observable, we have to take its expectation value.
Since we work in the Heisenberg picture as we have mentioned in Sec.~\ref{sec:expansion}, the operator is sandwiched by 
time-independent initial bra and ket. 
Let $| \Phi \rangle$ be such a ket vector which is constructed by multiplying the vacuum $| 0 \rangle$ by the creation operators 
at the initial time in an appropriate manner for a desired initial condition.
Then the electronic charge density $\rho_e(ct,\vec{r})$ is 
\begin{eqnarray}
\rho_e(ct,\vec{r})
=
\langle \Phi | :  \hat{\rho}_e(ct,\vec{r})  :  | \Phi \rangle
=
\sum_{p,q=1}^{N_D} \sum_{c,d=\pm} \rho_{p^c q^d}(\vec{r}) \langle \Phi | :  \hat{\cal E}_{p^c q^d}(t) :  | \Phi \rangle,
\label{eq:rhoe_ev}
\end{eqnarray}
where $: :$ denotes the operator in between should be normal ordered.
Alternatively, we can also say the physical quantity should be defined to have zero vacuum expectation value:
$\langle \Phi | :  \hat{\cal E}_{p^c q^d}(t) :  | \Phi \rangle =  \langle \Phi | \hat{\cal E}_{p^c q^d}(t) | \Phi  \rangle - \langle 0 | \hat{\cal E}_{p^c q^d}(t) | 0\rangle$.
This expectation value can be computed using the (anti-)commutation relations after $\hat{\cal E}_{p^c q^d}(t)$ is expressed by the operators at the initial time via the evolution equation Eq.~\eqref{eq:dcalEdt}.
To express $\hat{\cal E}_{p^c q^d}(t)$ in terms of the annihilation/creation operators at the initial time, we may discretize the time variable and solve Eqs.~\eqref{eq:dcalEdt} and \eqref{eq:dcalCdt} step by step in time direction. 
Although this procedure and subsequent computation of the expectation value can in principle be carried out, since we are dealing with the differential equations of operators, which are not commutative, the algebraic manipulation required is hugely demanding.
The difficulty would increase very rapidly as the time steps grow.
We consider such brute force computation is important to obtain as exact results as possible and it may not be impossible to do so employing recent developments in the field of computer algebra \cite{mathematica, symbolicc}.
At this stage, however, it is more important to look for a method to truck the time evolution of the physical quantities based on Eqs.~\eqref{eq:dcalEdt} and \eqref{eq:dcalCdt} with suitable approximations.
We will discuss such a method in next section.

\section{Approximation by density matrix equations}  \label{sec:approximation}

At the end of the previous section, we have pointed out that the excitation operator equations of the Rigged QED, 
Eqs.~\eqref{eq:dcalEdt} and \eqref{eq:dcalCdt}, are practically impossible to solve by the present computational technology. 
We therefore propose an approximation method, which is described in this section.  
We also show a result of numerical calculation based on that approximation for a hydrogen atom. 

\subsection{Density matrix equations} \label{sec:dm}

We begin by introducing density matrices for electrons and atomic nuclei.
They are defined by the expectation values (Sec.~\ref{sec:evolution_exphys}) of the corresponding excitation operators and we denote them as
${\cal E}_{p^c q^d}$ and ${\cal C}_{a ij}$ respectively. 
Namely, the electron density matrix ${\cal E}_{p^c q^d}$ is 
\begin{eqnarray}
{\cal E}_{p^c q^d}(t) \equiv \langle \Phi | :  \hat{\cal E}_{p^c q^d}(t) :  | \Phi \rangle,
\end{eqnarray}
and the atomic nucleus density matrix ${\cal C}_{a ij}$ is
\begin{eqnarray}
{\cal C}_{a ij}(t) \equiv \langle \Phi | : \hat{\cal C}_{a ij}(t) :   | \Phi \rangle.
\end{eqnarray}

The first step of our approximation is to replace $\hat{\vec{A}}_A$ and $\hat{\vec{j}}_T$ by their expectation values, which are denoted by
\begin{eqnarray}
 \vec{\cal A}_A  &\equiv&  \langle \Phi | : \hat{\vec{A}}_A : | \Phi \rangle, \\
 \vec{\cal J}_T  &\equiv&  \langle \Phi | : \hat{\vec{j}}_T : | \Phi \rangle.
\end{eqnarray}
Using the density matrices, they are expressed as
\begin{eqnarray}
{\cal J}_T^k(x) &=&  {\cal J}^k(x) - {\cal J}_L^k(x), \label{eq:JT_approx} \\
{\cal J}^k(x) &=&
\sum_{p,q=1}^{N_D} \sum_{c,d=\pm}  j^k_{p^c q^d}(\vec{r}) {\cal E}_{p^c q^d} \nonumber \\
& &+ \sum_{a=1}^{N_n} \sum_{i,j=1}^{N_S}  \left\{ j_{aij}^k(\vec{r}) {\cal C}_{aij}
-  \frac{(Z_a e)}{m_a c} \rho_{aij}(\vec{r}) \left( \langle \hat{A}_{\rm rad}^k \rangle {\cal C}_{aij} +  {\cal A}_A^k {\cal C}_{aij}  \right) \right\}, \label{eq:J_approx} \\
{\cal J}^k_L(x) 
&=& 
-\sum_{p,q=1}^{N_D} \sum_{c,d=\pm}  E^k_{p^c q^d}(\vec{r}) \frac{d {\cal E}_{p^c q^d}}{dt} 
-\sum_{a=1}^{N_n}\sum_{i,j=1}^{N_S}  E^k_{a ij}(\vec{r}) \frac{d {\cal C}_{a ij}}{dt},  \label{eq:JL_approx}
\end{eqnarray}
where $\langle \hat{A}_{\rm rad}^k \rangle = \langle \Phi | \hat{A}_{\rm rad}^k | \Phi \rangle $, and 
\begin{eqnarray}
\vec{\cal A}_A(ct, \vec{r}) = \frac{1}{c} \int d^3\vec{s}\, \frac{\vec{\cal J}_T(c u, \vec{s})}{|\vec{r}-\vec{s}|}, \qquad
u = t - \frac{|\vec{r}-\vec{s}|}{c}.   \label{eq:AA_approx}
\end{eqnarray}

The second step is to replace $\hat{A}_0$ by its expectation value. 
This leads to replacement of the excitation operators in $\hat{I}_{4 p^c q^d}$ and $\hat{I}_{4aij}$ 
(Eqs.~\eqref{eq:I4pq} and \eqref{eq:I4aij}) by corresponding density matrices. 

The final step is taking the expectation values of the time evolution equations of the excitation operators,
 Eqs.~\eqref{eq:dcalEdt} and \eqref{eq:dcalCdt}, to obtain those of the density matrices. 
Then, the evolution equation of electron density matrix is derived from Eq.~\eqref{eq:dcalEdt} as
\begin{eqnarray}
i\hbar \frac{d {\cal E}_{p^c q^d}}{dt} 
&=&
\sum_{r=1}^{N_D}  \sum_{e=\pm}
\left( -{\cal I}_{r^e p^c} {\cal E}_{r^e q^d}  + {\cal I}_{q^d r^e}  {\cal E}_{p^c r^e}  \right),
\label{eq:dcalEdt_dm}
\end{eqnarray}
where ${\cal I}_{p^c q^d}  \equiv {\cal I}_{1p^c q^d} + {\cal I}_{2p^c q^d}+ {\cal I}_{3p^c q^d} + {\cal I}_{4p^c q^d}$ and
\begin{eqnarray}
{\cal I}_{1p^c q^d} 
&=& T_{p^c q^d}, \\
 {\cal I}_{2p^c q^d} 
 &=& 
  -\frac{1}{c} \int d^3\vec{r}\, \vec{j}_{p^c q^d}(\vec{r}) \cdot \left(    \langle \hat{\vec{A}}_{\rm rad}(x) \rangle + {\vec{\cal A}}_A(x) \right), \\
  {\cal I}_{3p^c q^d} 
  &=& M_{p^c q^d}, \\
  {\cal I}_{4p^c q^d} 
  &=&
  \sum_{r,s=1}^{N_D} \sum_{e,f=\pm}   (p^c q^d | r^e s^f ) {\cal E}_{r^e s^f}
 + \sum_{a=1}^{N_n}\sum_{i,j=1}^{N_S} (p^c q^d | i_a j_a)  {\cal C}_{aij}.  
\end{eqnarray}
Similarly, the evolution equation of atomic nucleus density matrix is derived from Eq.~\eqref{eq:dcalCdt} as
\begin{eqnarray}
i\hbar \frac{d {\cal C}_{a ij}}{dt} 
&=&
\sum_{k=1}^{N_S} 
\left( - {\cal I}_{aki} {\cal C}_{a kj}  + {\cal I}_{ajk}  {\cal C}_{a ik}   \right)
\label{eq:dcalCdt_dm}
\end{eqnarray}
where we define ${\cal I}_{aij}  \equiv {\cal I}_{1aij} + {\cal I}_{2aij}+ {\cal I}_{3aij} + {\cal I}_{4aij}$ and 
\begin{eqnarray}
{\cal I}_{1aij} 
&=& T_{aij}, \\ 
{\cal I}_{2aij} 
&=&  -\frac{1}{c} \int d^3\vec{r}\, \vec{j}_{aij}(\vec{r}) \cdot \left( \langle \hat{\vec{A}}_{\rm rad}(x) \rangle +\vec{\cal A}_A(x) \right),\\
{\cal I}_{3aij} 
&=&  \frac{Z_a e}{2 m_a c^2} \int d^3\vec{r}\, 
\left(
\langle \hat{\vec{A}}_{\rm rad} \cdot \hat{\vec{A}}_{\rm rad} \rangle 
+ 2 \langle \hat{\vec{A}}_{\rm rad} \rangle \cdot \vec{\cal A}_A 
+ \vec{\cal A}_A \cdot \vec{\cal A}_A
\right)
\rho_{a ij}(\vec{r}),
\\
 {\cal I}_{4aij}
 &=& \sum_{p,q=1}^{N_D} \sum_{c,d=\pm} (p^c q^d | i_a j_a ) {\cal E}_{p^c q^d} 
 + \sum_{b=1}^{N_n} \sum_{k,l=1}^{N_S} (i_a j_a | k_b l_b)  {\cal C}_{bkl}.
\end{eqnarray}
Eqs.~\eqref{eq:dcalEdt_dm} and \eqref{eq:dcalCdt_dm}, together with 
Eqs.~\eqref{eq:JT_approx}-\eqref{eq:AA_approx},
form a closed set of time evolution equations of the density matrices. 
Since they are $c$-number quantities, they can be solved by a straightforward manner. 

\subsection{Numerical calculation}

We here show the result of numerical computation of solving the equations derived in Sec.~\ref{sec:dm}.
We adopt to make the equations simpler by employing further approximations.
One is neglecting $\vec{\cal A}_A$ contribution from the vector potential to avoid heavy numerical integration
involving the retarded potential. 
Another is the Born-Oppenheimer approximation (Sec.~\ref{sec:BO}) to eliminate the degree of freedom of the atomic nucleus density matrix.
Then, we need to only solve the equation for the electron density matrix Eq.~\eqref{eq:dcalEdt_dm} with 
${\cal I}_{p^c q^d}$  being 
\begin{eqnarray}
{\cal I}_{p^c q^d} =
h_{p^c q^d} + \sum_{r,s=1}^{N_D} \sum_{e,f=\pm}   (p^c q^d | r^e s^f ) {\cal E}_{r^e s^f}
  -\frac{1}{c} \int d^3\vec{r}\, \vec{j}_{p^c q^d}(\vec{r}) \cdot \langle \hat{\vec{A}}_{\rm rad}(x) \rangle,
 \label{eq:Ipq_app}
\end{eqnarray}
where the first term on the right-hand side is defined by Eq.~\eqref{eq:inth} and the third term can be 
written as Eq.~\eqref{eq:I2e_Arad} with $\hat{a}$ and $\hat{a}^\dagger$ replaced by 
$\langle \hat{a} \rangle$ and $\langle \hat{a}^\dagger \rangle$ respectively.

Now, we describe our computational setups for solving time evolution of electron charge density of a hydrogen atom.
The orbital functions which are used to expand the electron field operator (Sec.~\ref{sec:expansion_e}) are obtained 
by the publicly available DIRAC 10 code \cite{DIRAC10}. 
The Dirac-Coulomb Hamiltonian and the STO-3G basis set are used.  
Note that in this basis set, the electron density matrix is $4 \times 4$ matrix whose components denote
electron ($1^+$), its Kramers partner ($\bar{1}^+$), positron  ($1^-$) and its Kramers partner ($\bar{1}^-$).
The initial state for the electron is taken to be the ground state.
In terms of the electron density matrix, at the initial time, ${\cal E}_{1^+ 1^+} = 1$ and the other components are set to be zero.
As for the initial state for the photon, we perform calculation with and without initial photon field. 
When we include initial photon, we assume coherent state for the initial state so that 
$\langle \hat{a} \rangle$ and $\langle \hat{a}^\dagger \rangle$ have non-zero values. 
We take the coherent state to be that of a single mode with the eigenvalue equals to $1$.  
The mode is chosen to be in the $x$-direction ($\theta = \pi/2$ and $\phi=0$) and the polarization to be $\sigma = +1$
(see Sec.~\ref{sec:expansion_photon}). The momentum $p^0$ is taken to be 1, 10 and 20.
We work in the atomic units so that $m_e = e = \hbar = 1$ and $c = 137.035999679$.
The 1\,a.u.~of time corresponds to $2.419 \times 10^{-17}$\,s or $24.19$\,as.

\begin{figure}
\begin{center}
\includegraphics[width=12cm]{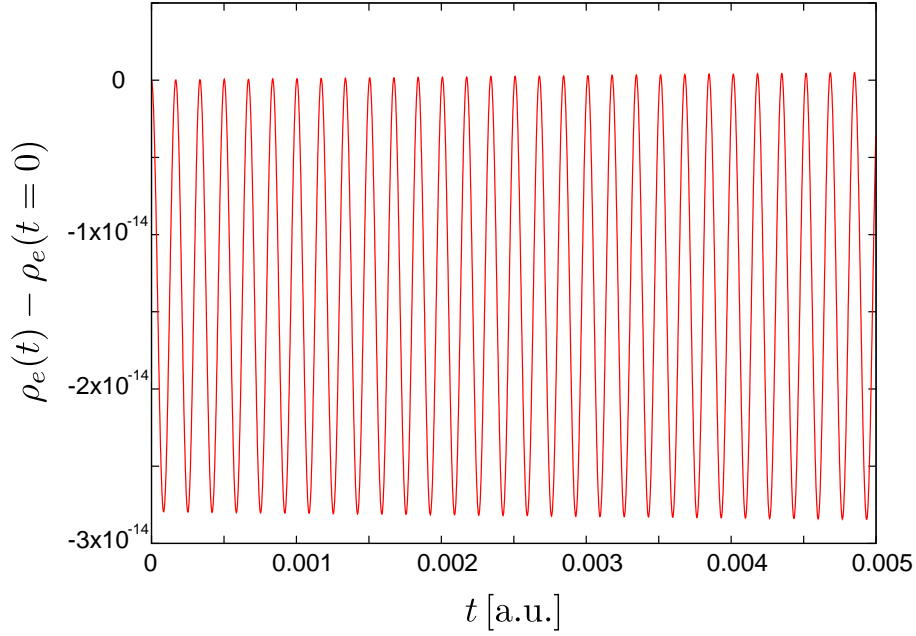}
\caption{The time evolution of charge density of the hydrogen atom at $(x, y, z) = (1, 0, 0)$. 
The variation from the initial value is plotted. The atomic units are used. 
There is no photon in the initial state.
}
\label{fig:plot_norad}
\end{center}
\end{figure}

\begin{figure}
\begin{center}
\includegraphics[width=12cm]{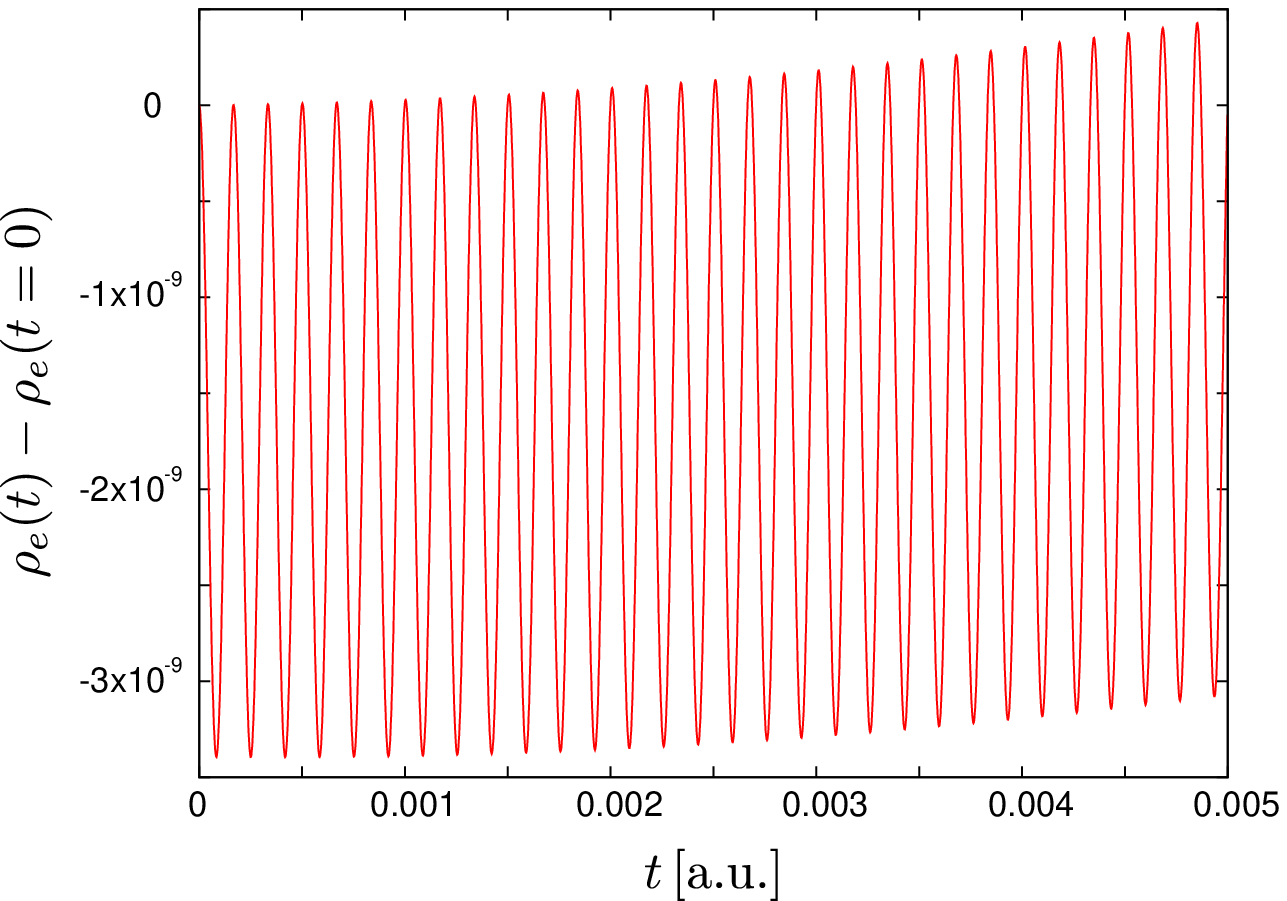}
\caption{Similar to Fig.~\ref{fig:plot_norad} but the initial photon state with $p^0 = 1$.}
\label{fig:plot_p2}
\end{center}
\end{figure}

\begin{figure}
\begin{center}
\includegraphics[width=12cm]{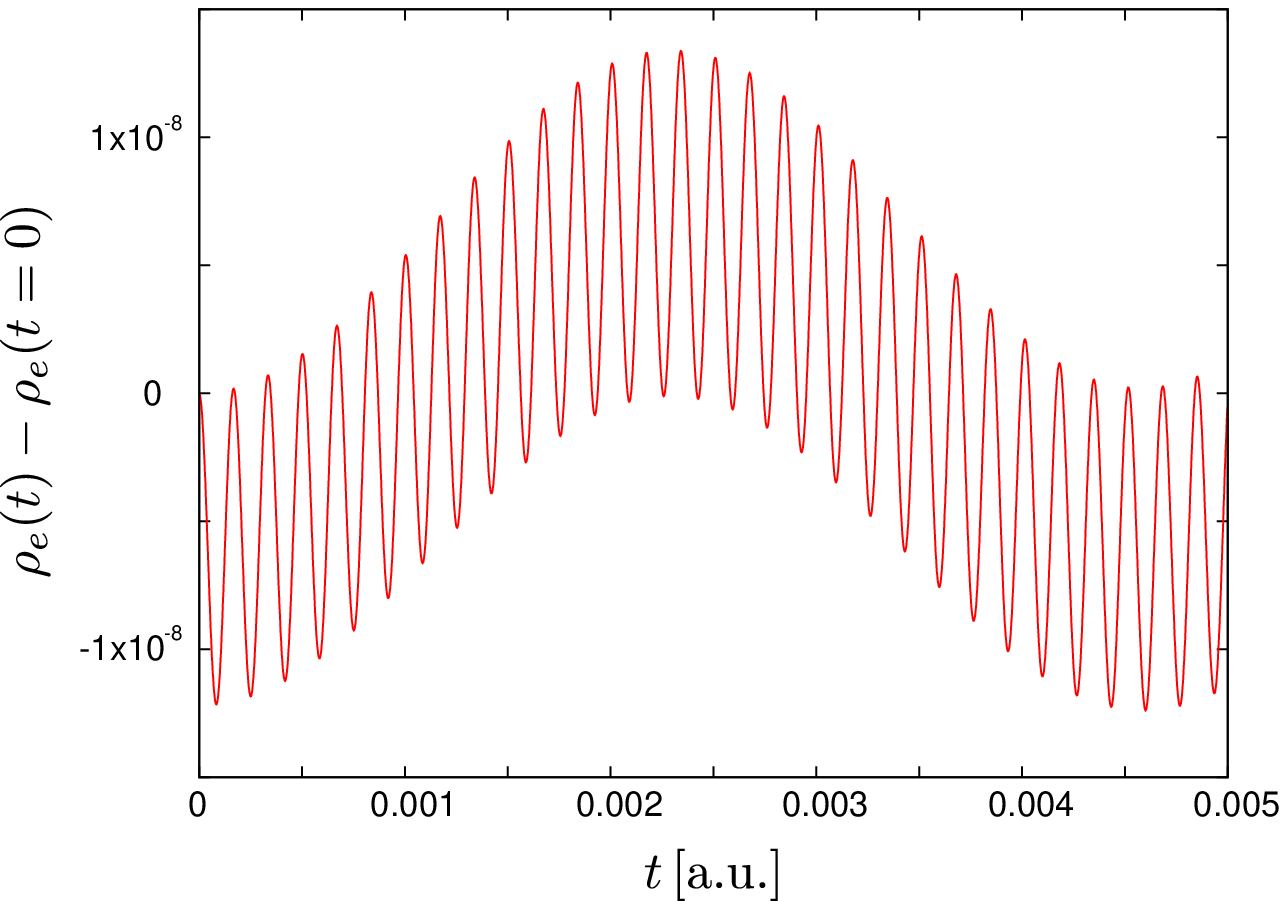}
\caption{Similar to Fig.~\ref{fig:plot_norad} but the initial photon state with $p^0 = 10$.}
\label{fig:plot_p20}
\end{center}
\end{figure}

\begin{figure}
\begin{center}
\includegraphics[width=12cm]{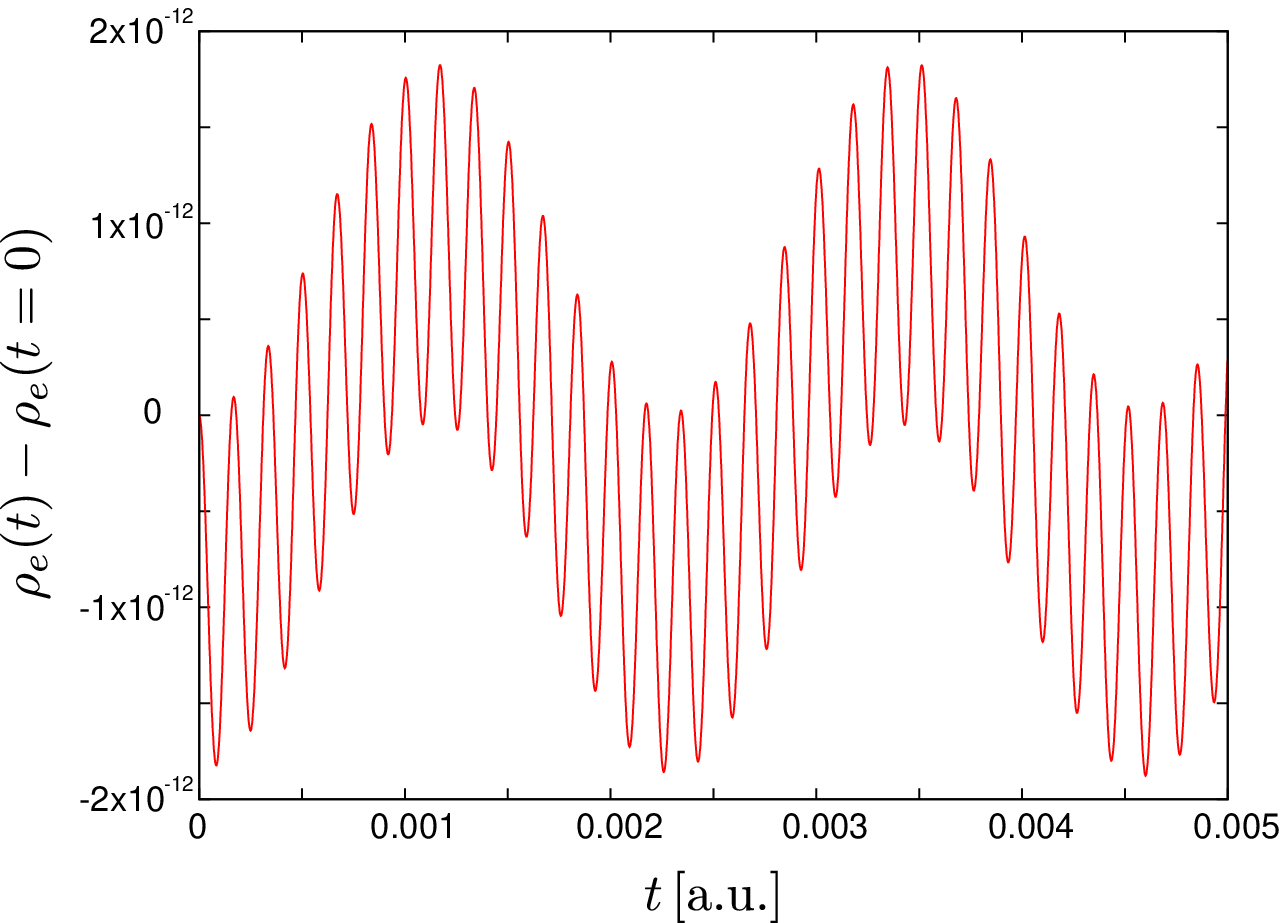}
\caption{Similar to Fig.~\ref{fig:plot_norad} but the initial photon state with $p^0 = 20$.}
\label{fig:plot_p40}
\end{center}
\end{figure}

As is described in Sec.~\ref{sec:evolution_exphys}, the electronic charge density $\rho_e(ct,\vec{r})$ is calculated as 
Eq.~\eqref{eq:rhoe_ev}.
In our approximation by the density matrix, 
\begin{eqnarray}
\rho_e(ct,\vec{r})
=
\sum_{p,q=1}^{N_D} \sum_{c,d=\pm} \rho_{p^c q^d}(\vec{r}) {\cal E}_{p^c q^d}(t).
\end{eqnarray}
We compute this quantity at $(x, y, z) = (1, 0, 0)$ and plot the time evolution in Figs.~\ref{fig:plot_norad}-\ref{fig:plot_p40}. 
Fig.~\ref{fig:plot_norad} shows the result with no photon in the initial bra and ket.
Figs.~\ref{fig:plot_p2}, \ref{fig:plot_p20} and \ref{fig:plot_p40} show the result with the initial photon 
bra and ket taken as a coherent state (as describe above in detail) with $p^0 = 1$, 10 and 20 respectively. 

\begin{figure}
\begin{center}
\includegraphics[width=12cm]{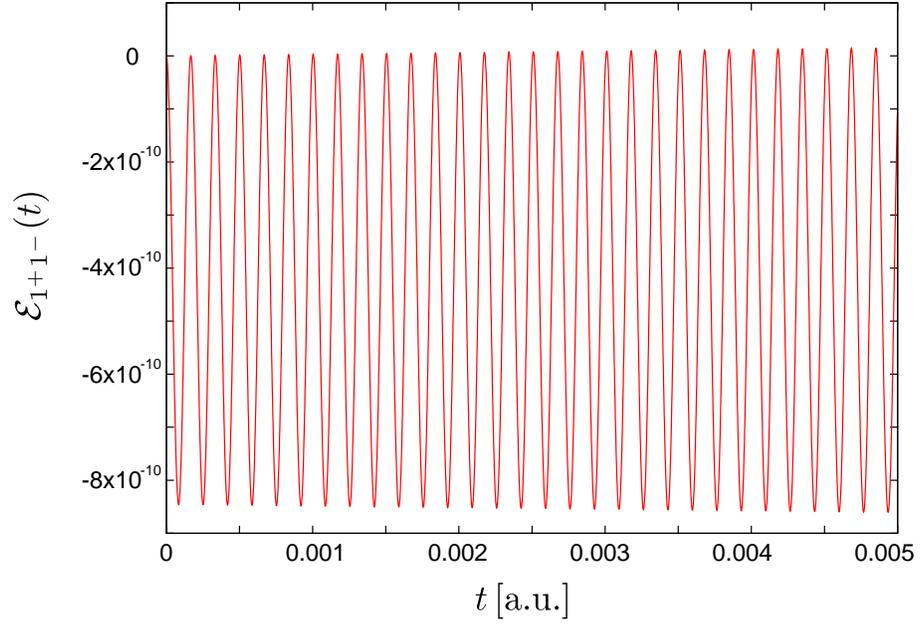}
\caption{The time evolution of the $(1^+, 1^-)$-component of the electron density matrix for no photon in the initial state.
}
\label{fig:plotdm13_norad}
\end{center}
\end{figure}

The common feature of these results is the oscillatory behavior. 
From the visual inspection of the figures, there are two types of oscillations. 
First, we see very rapid oscillations with the period of about $1.7 \times 10^{-4}$\,a.u. in all of Figs.~\ref{fig:plot_norad}-\ref{fig:plot_p40}. 
Second, as is seen in Figs.~\ref{fig:plot_p20} and \ref{fig:plot_p40}, there are oscillations with longer period 
which modulate the rapid oscillations. 
They have the period of roughly $4.6 \times 10^{-3}$\,a.u and $2.3 \times 10^{-3}$\,a.u in Figs.~\ref{fig:plot_p20} and \ref{fig:plot_p40} respectively. 
These numbers suggest the origins of the oscillations. 
The period of rapid ones is very close to the period which is determined from mass scales of electron and positron,
that is, $2\pi/(2m_e c^2) = 1.67 \times 10^{-4}$. 
The longer periods seen in Figs.~\ref{fig:plot_p20} and \ref{fig:plot_p40} are very close to the period which is 
determined from the initial photon momentum, $2\pi/ (p^0 c)$, giving $4.59 \times 10^{-3}$ and $2.29 \times 10^{-3}$
for $p^0=10$ and 20 respectively. 
The interpretation of the latter oscillations is that they are caused by the initial photon state which operates as the external 
oscillating electromagnetic field. 
Technically, they originate from the third term of Eq.~\eqref{eq:Ipq_app} and the time scale of the oscillations
is easily read off from Eq.~\eqref{eq:I2e_Arad}.
As for the former rapid oscillations, since it has the period of twice the mass of electron (or positron),
it can be interpreted as the fluctuations originated from virtual electron-positron pair creations. 
The numerical origin is of course the mass term, Eq.~\eqref{eq:intM}, in the first term of Eq.~\eqref{eq:Ipq_app}.
To see this electron-positron oscillations more explicitly, in Fig.~\ref{fig:plotdm13_norad}, we plot  the time evolution of
 $(1^+, 1^-)$-component of the electron density matrix in the case of no photon in the initial state.
 (Note that $1^+$ denotes electron and $1^-$ positron as explained earlier.)
We see the oscillations with the period same as those in Fig.~\ref{fig:plot_norad}.
Although some of the other components exhibit similar oscillatory behavior, their amplitudes are much smaller and 
almost all of the contribution to the oscillations come from the $(1^+, 1^-)$-component.

\section{Conclusion}  \label{sec:conclusion}

In this paper, we have discussed how we formulate time evolution of physical quantities in the framework of the Rigged QED
treating non-perturbatively the interactions among electrons, atomic nuclei and photons.
We have defined the time-dependent annihilation/creation operators for the electron and the atomic nucleus from
corresponding field operators by expanding them by appropriate functions of spatial coordinates. 
The photon fields have been shown to expressed by the photon annihilation/creation operators of 
the free photon field and those of the electron and the atomic nucleus. 
This enabled us to include the interactions mediated by the photon field in a non-perturbative manner. 
We then have derived the time evolution equations for the excitation operators of the electron and the atomic nucleus
and sketched how physical quantities can be computed from them.

We have pointed out that the last parts of the procedure are very computationally demanding.
Solving the coupled evolution equations for the electron and nucleus excitation operators 
by a finite difference method in time requires sophisticated coding technique due to
non-commutativity of the operators. 
The existence of the retarded potential makes the time evolution method more difficult. 
Also, since the coefficients of the equations involve spatial integration of up to six dimensions, 
their computation is very time-consuming. 
After that, similar difficulty of non-commutative algebra lies in computing expectation values of the time-evolved excitation operators
to obtain time-evolved physical quantities. 
This is somewhat alleviated by the use of the Wick's theorem but it still requires large computational 
resource even with small number of time steps.
It is true that such brute force evaluation can be carried out for a few time steps for a small system and
would be feasible in future for longer time scale and larger systems considering the recent development
in computer science and technology. 
Also, it is certainly meaningful to obtain accurate results in such a way. 
However, at this beginning stage, we consider it is more important to look for approximation methods to truck the time evolution of 
the physical quantities without too much computational time.

Therefore, we have proposed a method to approximate the time evolution equations of the operators by 
replacing some operators with their expectation value in the evolution equations. 
In other words, in an exact sense, we have to compute expectation value after the operators are evolved,
but, in our approximation, some operators are replaced by their expectation values and expectation values are evolved.
Although this method has room for improvement, we consider this is a good point to start. 
In fact, the numerical result of the time evolution of charge density of a hydrogen atom exhibits the oscillatory feature
which is considered to originate from the electron-positron pair creations. 
This shows that our simplified way of computation can reproduce one of the most notable features of QED.
Even within the framework of this approximation, there are a lot more works to be done. 
In our near future work, we will remove the Born-Oppenheimer approximation and will include the full effect of 
vector potential.



\appendix

\section{Born-Oppenheimer approximation}  \label{sec:BO}

In this section, we show the time evolution equation of the operators for the electron under 
the Born-Oppenheimer (BO) approximation. Namely, we fix positions of the nuclei. 
Since the Rigged QED is proposed to include nucleus motion, this approximation is 
somewhat contradictory. 
However, there are many phenomena which can be described 
under the BO approximation.
Moreover, since it eliminates the nucleus degree of freedom, the equations become
extremely simpler and more accurate results can be obtained with less difficulty
compared with the non-BO case. 
Thus, we consider it would be useful to present here the BO approximated version of the
evolution equations. 

Under the BO approximation, the atomic nuclear charge density operator $\hat{\rho}_a(x)$ (Eq.~\eqref{eq:rhoa})
is given by 
\begin{eqnarray}
\hat{\rho}_a(x) = Z_a e\, \delta(\vec{r} - \vec{R}_a),
\end{eqnarray}
where $\vec{R}_a$ is the position for the nucleus $a$, 
$\vec{R}_a = \vec{R}_{a_1} \oplus \vec{R}_{a_2} \oplus \cdots \oplus \vec{R}_{a_{n_a}}$
specified by the generic direct sum of $c$-numbered $n_a$ vectors that are clamped in space if any, 
with which mandatory manipulation for a function $f$ of $\vec{R}_a$ should read 
$f(\vec{R}_a) = \sum_{k=1}^{n_a} f(\vec{R}_{a_k}) $ as made obvious, 
and  the atomic nuclear charge current density operator $\hat{\vec{j}}_a(x)=\vec{0}$.
 
This approximates the operators shown in \ref{sec:charge} as follows. 
Eq.~\eqref{eq:A0_2} is simplified as
\begin{eqnarray}
\hat{A}_0(ct, \vec{r})
 =
  \sum_{p,q=1}^{N_D} \sum_{c,d=\pm} V_{p^c q^d}(\vec{r}) \hat{\cal E}_{p^c q^d}
 + \sum_{a=1}^{N_n} \frac{Z_a e}{|\vec{r} - \vec{R}_a|},   \label{eq:A0_2_BO}
\end{eqnarray}
Eq.~\eqref{eq:jtot2} as
 \begin{eqnarray}
\hat{j}^k(x) &=&
\sum_{p,q=1}^{N_D} \sum_{c,d=\pm}  j^k_{p^c q^d}(\vec{r}) \hat{\cal E}_{p^c q^d},
 \label{eq:jtot2_BO}
\end{eqnarray}
and Eq.~\eqref{eq:jL} as
\begin{eqnarray}
\hat{j}^k_L(x) 
&=& 
-\sum_{p,q=1}^{N_D} \sum_{c,d=\pm}  E^k_{p^c q^d}(\vec{r}) \frac{d \hat{\cal E}_{p^c q^d}}{dt}.
  \label{eq:jL_BO}
\end{eqnarray}
Then  $\hat{\vec{A}}_A(ct, \vec{r})$ can be written as
\begin{eqnarray}
\hat{\vec{A}}_A(ct, \vec{r})
&=& 
  \frac{1}{c} \sum_{p,q=1}^{N_D} \sum_{c,d=\pm} \int d^3\vec{s}\, 
\left\{ 
\frac{\vec{j}_{p^c q^d}(\vec{s})}{|\vec{r}-\vec{s}|}  \hat{\cal E}_{p^c q^d}(u)
+   \frac{\vec{E}_{p^c q^d}(\vec{s})}{|\vec{r}-\vec{s}|}  \frac{d \hat{\cal E}_{p^c q^d}}{dt}(u) 
\right\},  
\end{eqnarray}
where $u = t - \frac{|\vec{r}-\vec{s}|}{c}$.

In addition, $\hat{I}_{4 p^c q^d}$  (Eq.~\eqref{eq:I4pq}) is simplified as
\begin{eqnarray}
\hat{I}_{4 p^c q^d} 
&=&
   \sum_{r,s=1}^{N_D} \sum_{e,f=\pm}   (p^c q^d | r^e s^f ) \hat{\cal E}_{r^e s^f}
 + \sum_{a=1}^{N_n} (Z_a e) V_{p^c q^d}(\vec{R}_a).
\end{eqnarray}

Therefore, the BO approximation version of Eq.~\eqref{eq:dedt2} can be written as
\begin{eqnarray}
i\hbar \frac{\partial \hat{e}_{p^c}}{\partial t} 
&=&
\sum_{q=1}^{N_D}  \sum_{d=\pm} h_{p^c q^d} \hat{e}_{q^d}
+
\sum_{q,r,s=1}^{N_D}  \sum_{d,e,f=\pm}  (p^c q^d | r^e s^f ) \hat{\cal E}_{r^e s^f} \hat{e}_{q^d} \nonumber \\
& &\hspace{-1.5cm} -\frac{1}{c^2}  \sum_{q,r,s=1}^{N_D}  \sum_{d,e,f=\pm} \int d^3\vec{r}\,  d^3\vec{s}\, \left\{ 
\frac{\vec{j}_{p^c q^d}(\vec{r}) \cdot \vec{j}_{r^e s^f}(\vec{s})}{|\vec{r}-\vec{s}|}  \hat{\cal E}_{r^e s^f}(u)
+   \frac{\vec{j}_{p^c q^d}(\vec{r}) \cdot \vec{E}_{r^e s^f}(\vec{s})}{|\vec{r}-\vec{s}|}  \frac{d \hat{\cal E}_{r^e s^f}}{dt}(u) 
\right\}
   \hat{e}_{q^d} \nonumber \\
& &\hspace{-1.5cm}-\frac{1}{c} \frac{\sqrt{4\pi \hbar^2 c}}{\sqrt{(2\pi\hbar)^3}}
\sum_{q=1}^{N_D}  \sum_{d=\pm}  \sum_{\sigma=\pm1} \int \frac{d^3 \vec{p}}{\sqrt{2p^0}} \times \nonumber \\
& &\hspace{-0.5cm} \left[ 
\vec{F}_{p^c q^d}(\vec{p}) \cdot \vec{e}(\vec{p},\sigma) e^{-i c p^0 t/\hbar} \hat{a}(\vec{p},\sigma)\hat{e}_{q^d}
+\vec{F}_{p^c q^d}(-\vec{p}) \cdot \vec{e^*}(\vec{p},\sigma) e^{i c p^0 t/\hbar}   \hat{a}^\dagger(\vec{p},\sigma)   \hat{e}_{q^d}
\right],
\label{eq:dedt2_BO}
\end{eqnarray}
where we define
\begin{eqnarray}
h_{p^c q^d} \equiv T_{p^c q^d}+ M_{p^c q^d} + \sum_{a=1}^{N_n} (Z_a e) V_{p^c q^d}(\vec{R}_a). \label{eq:inth}
\end{eqnarray}
Under the BO approximation, this is the only equation which governs the time evolution of the system.


\newpage

\end{document}